\documentclass[aps,prd,twocolumn,showpacs,superscriptaddress,groupedaddress,nofootinbib]{revtex4}  
\usepackage{graphicx}  
\usepackage{dcolumn}   
\usepackage{bm}        
\usepackage{amssymb}   
\usepackage{amsmath}
\usepackage{soul}

\RequirePackage[colorlinks,citecolor=blue,urlcolor=black,linkcolor=blue]{hyperref}

\RequirePackage{graphicx}

\begin{document}

\title{ SKA-Phase 1 sensitivity to synchrotron radio emission from multi-TeV Dark Matter candidates}
\author{Jose A. R. Cembranos}
\affiliation{Departamento de F\'isica Te\'orica I and IPARCOS, Universidad Complutense de Madrid, E-28040 Madrid, Spain}
\author{\'Alvaro de la Cruz-Dombriz}
\affiliation{Cosmology and Gravity Group, Department of Mathematics and Applied Mathematics, University of Cape Town, Rondebosch 7701, Cape Town, South Africa}
\author{Viviana Gammaldi}
\affiliation{Instituto de F\'isica Te\'orica (IFT UAM-CSIC), calle N. Cabrera 13-15, 28049, Madrid, Spain}
\affiliation{Universidad Aut\'onoma Madrid (UAM), Campus Cantoblanco, 28049, Madrid, Spain}
\affiliation{Scuola Internazionale Superiore di Studi Avanzati (SISSA), Trieste, Italy}
\affiliation{Istituto Nazionale di Fisica Nucleare (INFN)}
\author{Miguel M\'endez-Isla}
\affiliation{Cosmology and Gravity Group, Department of Mathematics and Applied Mathematics, University of Cape Town, Rondebosch 7701, Cape Town, South Africa}       
\date{\today}

\begin{abstract}
In the era of radio astronomy, the high sensitivity of the Square Kilometre Array (SKA) could play a decisive role in the detection of new radio sources. In this work, we study the SKA sensitivity to the synchrotron radio emission expected by the annihilation of TeV DM candidate in the Draco dwarf spheroidal galaxy. On the one hand, we consider model-independent DM candidates: we find out that with 1000 hours of data-taking, SKA1-MID will be able to exclude up to 10 TeV thermal DM candidates that annihilate in $W^+W^-$ and $b\bar b$ channels.  We also study as these constraints improve by including a density enhancement due to a DM-spike associated with an intermediate-mass black hole in Draco. On the other hand, we consider extra-dimensional brane-world DM candidates, dubbed branons. In this specific scenario, SKA allows us to set constraints on the branon parameter space ($f$,\,$M$), where $f$ is related to the coupling of the branon to the Standard Model particles and $M$ is the mass of the branon itself. In particular, we consider two different branon DM candidates. We find out that SKA will be able to set more stringent constraints on the branon DM candidate required in order to fit the AMS-02 data, yet the sensitivity of the instrument should be improved in order to study the branon candidate for the Galactic Centre. Nonetheless, we show that SKA represents  - among other detectors - the most promising instrument for multi-wavelength detection of synchrotron radio emission by annihilating multi-TeV DM.
\end{abstract}

\maketitle

\section{Introduction}
\label{sec:intro}

Dark Matter (DM) is a fundamental ingredient to explain several observations in our Universe at both astrophysical and cosmological scales  \cite{clowe2004weak,clowe2006direct,rubin1970rotation,rubin1980rotational,massey2010dark,anderson2014clustering}, yet its nature still remains elusive \cite{Zwicky, Planck1, Planck2, Bergstrom:2000pn, Bertone:2004pz}. Indirect DM searches conform to different detection strategies to disentangle the main features of DM, e.g. multimessenger and multi-wavelength approaches. 
%
In the first approach, DM particles may annihilate in galactic halos and produce cosmic rays that can be detected by observatories of a different kind. 
Nonetheless, the interaction of charged cosmic rays with the environment (i.e. galactic magnetic field)
 and their energy loss would generate 
 photons in a large range of frequencies.
%
%
The study of these signals sets the basis of DM indirect searches through multiwavelength astronomy; depending on the energy loss mechanism distinctive signals could be detected at different frequencies, either via synchrotron emission, Inverse Compton Scattering (ICS), bremsstrahlung or Coulombian interactions. 
In particular, synchrotron emission from secondary electrons and positrons produced by annihilations of Weakly Interacting Massive Particles (WIMPs) could be detected at radio frequencies with the  Square Kilometre Array (SKA) \cite{Dewdney:2013ard}.
The SKA-Phase $1$ (SKA1) is claimed to be one of the most promising instruments in radio astronomy. Its  high sensitivity and resolution could be key to solve open questions in both Astrophysics, Cosmology and Particle Physics, including Physics beyond the Standard Model (SM) ({\it c. f.} \cite{Dewdney:2013ard} for further details).\\

Radio emission from GeV DM candidates has been already investigated in numerous studies \cite{Spekkens:2013ik, Natarajan:2015hma, Colafrancesco:2006he, McDaniel:2017ppt, Colafrancesco:2005ji}. In this work, we will focus on radio emission expected by the annihilation of TeV DM particles. In particular, we study the SKA sensitivity to prospect radio signals that could be produced by synchrotron emission due to secondary 
fluxes of $e^+/e^-$ \cite{Feng:2017tnz,Cheng:2016slx,Carquin:2015uma,Delahaye:2007fr, 
Feng:2013zca, Ibarra:2013zia} after the annihilation of TeV DM particles. 
Among other DM candidates, it has been argued that SKA1 would be able to detect 
radio synchrotron emission from minimal-supersymmetric (MSSM) DM candidates up to masses of tens TeV 
\cite{Kar:2018rlm}. Nonetheless, only a limited number of theories could naturally produce DM particles with masses heavier than tens TeV ({\it c. f.} dark atoms \cite{Belotsky:2014haa,Belotsky:2016tja}, minimal DM models \cite{Cirelli:2009uv,Garcia-Cely:2015dda} or fermionic DM \cite{Chua:2013zpa}). In this regard, brane-world theories may also naturally produce thermal DM candidates up to masses of 100 TeV \cite{Cembranos:2003mr}.

In this work, we will focus on the Draco dwarf spheroidal (dSph) galaxy, i.e. a benchmark target for DM searches. 
Due to the low percentage of baryonic matter with respect to the DM abundance, also the radio background component - due to the interaction of cosmic rays of astrophysical origin with the environment -  is expected to be subdominant in dSph. 
In fact, dSph are 
faint radio sources requiring detectors with high sensitivity. 
For instance, no detection limits of several dSphs with the Australia Telescope Compact Array (ATCA) \cite{Regis:2014joa} and the Green Bank Telescope (GBT) \cite{Spekkens:2013ik,Natarajan:2015hma}, have already achieved sensitivities as low as 
the order of  mJy  and the SKA
would be able to reach $\mu$Jy orders of magnitude \cite{Dewdney:2013ard}. 
\\

The manuscript is organised as follows: in Section \ref{sec:Sync} we introduce  
the physics of diffusion
of charged particles in the galaxy 
and the generalities of DM indirect detection by synchrotron emission. 
In Section \ref{sec:SKA} we 
briefly characterise the most relevant SKA1 technicalities. 
In Section \ref{sec:Sensitivity} we discuss the SKA1 sensitivity to synchrotron radio emission from TeV DM candidates. Firstly, we adopt a model-independent approach and we consider prospective boosts in the DM distribution due to a Black Hole (BH)-related DM spike. Secondly, we will consider 
brane-world theories and branons as prospective TeV WIMPs. 
In Section \ref{sec:Discussion} we will briefly discuss the possibility to detect multiwavelength emission from TeV DM candidates with current and future generation of telescopes.
Finally, Section \ref{sec:Conclusions}
is dedicated to draw the main conclusions of this study.    

\section{Synchrotron emission}
\label{sec:Sync}


During
the transport of cosmic rays in the galactic environment, 
the deflection of charged particles by the galactic magnetic field would result in the emission of electromagnetic radiation. Contingent upon the velocity regime of those particles, two distinct emissions could be detected.
On the one hand, in the case of non-relativistic particles, 
emission lines are associated with the cyclotron frequency with a two-lobe distribution around the direction of acceleration. 
 In this regard, only some environments with high magnetic fields, such as neutron stars, would render 
this kind of radiation detectable.
%
On the other hand, in the case of ultra-relativistic particles, the emission is  produced through synchrotron radiation in a continuous frequency range. Indeed, ultra-relativistic $e^+/e^-$ are responsible for a large number of signatures in the sky, being the synchrotron emission one of the main mechanisms of energy losses. 
Such an $e^+/e^-$ propagation is  dominated by the diffusion equation 
\begin{eqnarray}
-\nabla\cdot\left[D\left(\textbf{r},E\right)\nabla\psi\right]-\frac{\partial}{\partial E}\left[b(\textbf{r},E)\psi\right]=Q_{e}(\textbf{r},E)\,,
\label{Diff_Eq_positrons}
\end{eqnarray} 
where $\psi\left(\textbf{r},E\right)$ is the number density of $e^+/e^-$ in equilibrium per unit of energy at the point $\textbf{r}$ of the galaxy, i.e., $\psi$ represents the spectrum of products after the propagation. $b(\textbf{r},E)$ represents the energy loss term and $D\left(\textbf{r},E\right)$ holds for the diffusion coefficient, which becomes $D\left(\textbf{r},E\right) \sim D\left(E\right)=D_{0} E^{\delta}$. In the following, we shall use $\delta=1/3$ according to the Kolmogorov description. It is important to highlight that the diffusion Eq. (\ref{Diff_Eq_positrons}) portrays a simplification of the Ginzburg-Syrovatsky transport equation. The latter takes into consideration some other mechanisms such as re-acceleration of cosmic rays (negligible in the case of ultra-relativistic $e^+/e^-$), spallation of cosmic rays, radioactive decay of nuclei of the ISM as well as eventual interactions with the galactic wind \cite{Strong:2007nh,Yuan:2017ozr}. However, all these mechanisms turn out to be negligible in dSph galaxies\footnote{The mass-to-light ratio in the case of Draco is $M/L \sim 300 M_{\odot}/L_{\odot}$ \cite{Lokas:2004sw}.}. 
\\

Ultra-relativistic $e^+/e^-$ are expected to be produced as a consequence of the annihilation of DM particles. Thus, the source term $Q_{e}(\textbf{r},E)$ in Eq. (\ref{Diff_Eq_positrons}) becomes:
\begin{eqnarray}
 Q_{e}(\textbf{r},E)=\frac{1}{2}\left\langle \sigma v\right\rangle\left(\frac{\rho_{\text{DM}}(\textbf{r})}{M}\right)^2\sum_{j}\beta_{j}\frac{{\rm d}N_e^j}{{\rm d}E}\,,
\label{Q}
\end{eqnarray}
which describes how DM particles with a thermally averaged cross section $\langle \sigma v \rangle$  annihilate and inject $e^+/e^-$ to the environment with a characteristic injection spectrum $\frac{{\rm d}N_e^j}{{\rm d}E}$ that depends on the $j$  annihilation channel. 
Once DM particles annihilate with a probability $\beta_{j}$ into a specific SM channel $j$, these SM particles could either decay or hadronise into cosmic rays (in our case $e^+/e^-$) that are injected into the galactic medium. In this work, the injection spectra $\frac{{\rm d}N_e^j}{{\rm d}E}$ have been computed using the functions provided by the software ${\rm PPC4DMID}$ \cite{Cirelli:2010xx},  including the electroweak effect that
becomes relevant for multi-TeV energy events \cite{Ciafaloni:2010ti}.
Such spectra establish, together with $\langle \sigma v \rangle$, $\beta_{j}$ and the mass of DM, $M$, the Particle Physics description for the DM model under study.  In the source term (\ref{Q}),  the DM density profile $\rho_\text{DM}(\textbf{r})$ takes into account the spatial distribution of DM in a particular target.
In the following, we shall focus on the isolated emission from the Draco dSph galaxy. 
%
For Draco dSph description, we adopt a Navarro-Frenk-White (NFW) DM density profile \cite{Colafrancesco:2006he}\footnote{In this work we considered a NFW profile, $\rho_{{\rm NFW}}(r)=\frac{\rho_s}{\frac{r}{r_s}\left(1+\frac{r}{r_s}\right)^2}$,  with $\rho_s=1.40$ $\text{GeV}/\text{cm}^{3}$ and $r_{s}=1$ $\text{kpc}$.}, although more sophisticated halo descriptions, such as the Kazantzidis profile \cite{kazantzidis2004density}, have also been applied to this specific galaxy \cite{sanchez2007dark,sanchez3530dark}. 

After a suitable change of variables, the diffusion equation, Eq. (\ref{Diff_Eq_positrons}), can be rewritten as a heat-like equation \cite{Colafrancesco:2005ji}, and thus, an analytical solution can be obtained in terms of a Green's function $G\left(r,E,E_{s}\right)$, i.e. 
\begin{eqnarray}
\psi(\textbf{r},E)=\frac{1}{b(\textbf{r},E)} \int^{M}_{E}{\rm d}E_{s}\,G\left(r,E,E_{s}\right) Q_{e}(\textbf{r},E).
\label{Sol_dif}
\end{eqnarray}
The spherical symmetry of the problem and its boundary conditions suggest the use of the image charge method applied to the Green's function, as explained in \cite{Baltz:1998xv,Colafrancesco:2005ji}. Such a method considers charges positioned at $r_{n}=(-1)^{n}r+2nr_{h}$, where the radius of diffusion $r_{h} $ is around $\sim2.5$ kpc for dSphs similar to Draco \cite{Colafrancesco:2006he}.
Thus $G\left(r,E,E_{s}\right)$ becomes 
\begin{eqnarray}
&&G\left(r,E,E_{s}\right)=\frac{1}{\sqrt{\pi \lambda_{D}^{2}(E,E_{s})}}\sum^{\infty}_{n=-\infty} \left(-1\right)^{n} \int^{r_{h}}_{0} {\rm d}r' \frac{r'}{r_{n}} \nonumber\\
&& \times \left(\frac{\rho_{\text{DM}}(r')}{\rho_{\text{DM}}(r)}\right)^{2} \left[  
\exp(-g_{n}^{-})-\exp(-g_{n}^{+})
\right],
\label{green}
\end{eqnarray}
with $g_{n}^{\pm}(r',E,E_{s})=\frac{(r' \pm r_{n} r)^{2}}{\lambda_{D}^{2}(E,E_{s})}$. 
 In the above expression, $\lambda_{D}(E,E_{s})$ represents the mean free path of $e^+/e^-$ from the coordinates of injection $(\textbf{r}_{s}, t_{s})$ with kinetic energy $E_{s}$ to a coordinate (\textbf{r}, $t$) with kinetic energy $E$ and satisfies 
%
%
\begin{eqnarray}
\lambda_{D}^{2}(E,E_{s})=4\int^{E_s}_{E}{\rm d}\varepsilon \frac{D(\varepsilon)}{b(\bf{r},\varepsilon)}\,\,,
\label{mean_free}
\end{eqnarray}
showing that the mean free path depends on both the diffusion $D(E)$ of cosmic rays in the medium,  and the 
loss of energy mechanism encapsulated in $b({\rm {\bf r}},E)$.  The latter quantity comprises
electromagnetic interactions affecting the  $e^+/e^-$ propagation 
 via bremsstrahlung, Coulombian, ICS and synchrotron emission 
and is defined as \cite{Sarazin:1999nz,Rybicki,delahaye2010galactic}
\begin{eqnarray}
b(\textbf{r},E)=b_{{\rm brem}}(E)+b_{{\rm Coul}}(E)+b_{{\rm ICS}}(E)+b_{{\rm syn}}(\textbf{r},E).
\label{b(E)_contributions}
\end{eqnarray}
\begin{figure*}
\centering
    \includegraphics[width=0.7\textwidth]{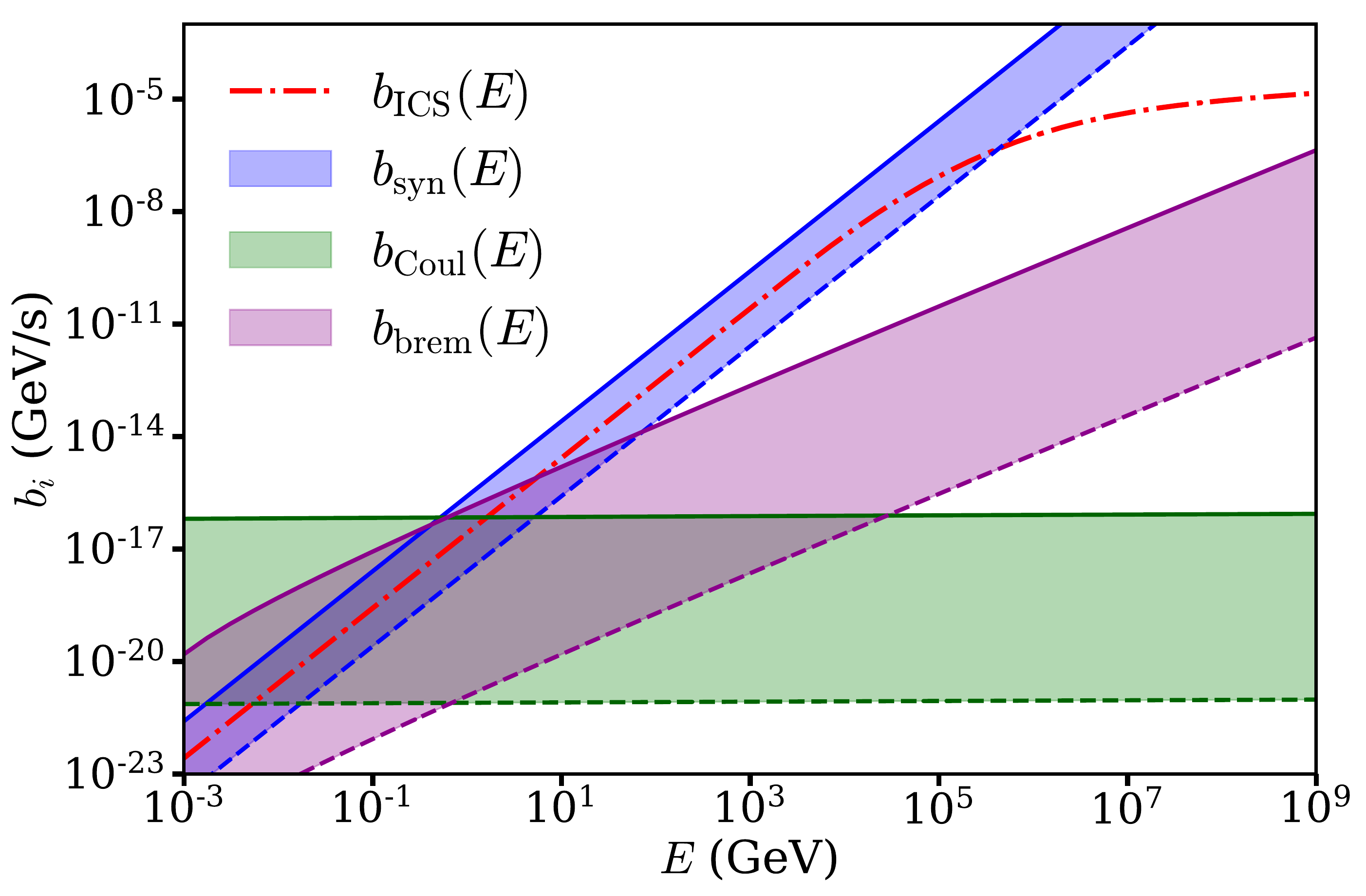}
\caption{Radiative losses $b_{i}$ dependence on the $e^+/e^-$ kinetic energy $E$. The blue band corresponding to synchrotron emission has been computed for a magnetic field between $B=1$ $\mu\text{G}$ (dashed line) and $B=10$ $\mu\text{G}$ (solid line). The bands for bremsstrahlung and Coulombian losses take into account the electron number density of the medium, $n_{e}$, between $10^{-6}$ $\text{cm}^{-3}$ (dashed) for dSphs and $0.1$  $\text{cm}^{-3}$ (solid) for galaxies. The analytical expressions for all the radiative losses in terms of $n_{e}$ and $B$ can be found following \cite{Sarazin:1999nz,Rybicki,McDaniel:2017ppt}. While Coulombian losses dominate at low energies, synchrotron emission and ICS lead the $e^+/e^-$ emission signals at higher energies. For ICS losses both the Thomson and Klein-Nishina regimes have been considered in the interaction of $e^+/e^-$ with the CMB photons with a temperature $T_{0}=2.73$ K \cite{delahaye2010galactic}.}
\label{Losses}
\end{figure*}
Depending on the astrophysical environment, some of the above mechanisms prevail over others at different energies as illustrated in Figure \ref{Losses}, where energy losses due to 
bremsstrahlung, Coulombian, ICS and synchrotron emission have been calculated\footnote{$b_{{\rm brem}}(E)=1.51\cdot 10^{-16}n_{e}E\left[\log(E/m_{e}) + 0.36\right],  
\\
b_{{\rm syn}}(\textbf{r},E)= 0.0254\cdot 10^{-4} B^{2}(\textbf{r}) E^{2},\\ 
b_{{\rm Coul}}(E)=6.13\cdot 10^{-16}n_{e}\left[1 + \log(E/n_{e}m_{e})/75\right],\\ 
b_{{\rm  ICS}}(E)=0.25\cdot 10^{-16}E^{2}\,\,\text{if}\,\,\frac{\gamma k_{b} T_{0}}{m_{e}c^{2}}<3.8\cdot10^{-4}, \\
b_{{\rm  ICS}}(E)=\frac{E^{2}m_{e} c^{2}\left(k_{b} T_{0}\right)^{3}}{\gamma}\exp\left[\sum^{5}_{i=0}c_{i}\left(\ln\frac{\gamma k_{b} T_{0}}{m_{e}c^{2}}\right)^{i}\right]  \,\,\text{if}\,\,3.8\cdot10^{-4}<\frac{\gamma k_{b} T_{0}}{m_{e}c^{2}}<1.8\cdot10^{3},\\
 b_{{\rm  ICS}}(E)=\frac{\sigma_{T}}{16}\frac{\left(m_{e}k_{b} T_{0}\right)^{2}}{\hbar^{3}}\left(\ln \frac{4 \gamma k_{b} T_{0}}{m_{e}c^{2}}-1.9805\right)   \,\,\text{if}\,\,1.8\cdot10^{3}<\frac{\gamma k_{b} T_{0}}{m_{e}c^{2}},$\\
with $c_{i}$ taken from \cite{delahaye2010galactic}. The constants $\sigma_{T}, k_{b}$ and $T_{0}$  are the Thomson cross section, the Boltzmann constant and the temperature of the gas of photons respectively.}.
There, colour bands  allow us to visualise the dependence of the energy losses both with the magnetic field (between $1$ $\mu$G and $10 $ $\mu$G) and the electron density of the medium, that we assume  $n_{e}=10^{-6}\text{cm}^{-3}$ \cite{Colafrancesco:2006he} for Draco. 
Figure \ref{Losses} also shows that synchrotron and ICS are predominant at TeV - and higher - scales, whereas bremsstrahlung and Coulombian losses dominate the sub-GeV range of energies. This thorough analysis of predominances in Eq. (\ref{b(E)_contributions}) turns out to be useful to accelerate the numerical determination of $\psi$ in Eq.  (\ref{Sol_dif}).
Indeed, since the exponentials in Eq. (\ref{green}) mix spatial coordinates with energy, the computation of integrals in Eq. (\ref{green}) might become difficult. 
Following the usual approach in \cite{Colafrancesco:2006he,McDaniel:2017ppt}, in order to compute the Green's function in Eq. (\ref{green}) we shall approximate $b_{{\rm syn}}(\textbf{r},E)\sim b_{{\rm syn}}(E)= 2.54\cdot 10^{-6} B_{\text{avg}}^{2} E^{2}$, with an averaged magnetic field $B_{\text{avg}}= 0.5\, \mu\text{G}$. This approximation is solely considered in the determination of Eq. (\ref{mean_free}), being the spatial dependence in $b_{\rm syn}(\textbf{r},E)$ fully retaken whenever Eqs. (\ref{Diff_Eq_positrons}) and (\ref{Sol_dif}) are considered. Indeed, as a result of the interaction of $e^+/e^-$ with the magnetic field, the predominant emission of secondary photons at radio frequencies is expected to happen through synchrotron emission. 
 In more detail, each $e^+/e^-$ produces a radiative emission. The emitted power of a single $e^+/e^-$ at redshift $z\approx 0$ is

\begin{eqnarray}
P_\text{syn}(\nu,\textbf{r}, E)&=& \int^{\pi}_{0} {\rm d}\alpha \frac{{\rm \sin}^{2}\alpha}{4\pi \epsilon_{0}} \frac{\sqrt{3}e^{3} B(\textbf{r})}{m_{e} c} F_{i} \left( \frac{\nu}{\nu_{c}(\textbf{r},E) {\rm \sin}\alpha}\right)
\label{P_syn},\nonumber\\
&&
\end{eqnarray}
where
\begin{eqnarray}
F_{i}(s)&=&s \int^{\infty}_{s} {\rm d}\xi\, K_{\frac{5}{3}}(\xi) \simeq \frac{5}{4} s^{\frac{1}{3}} \exp(-s) \left(648+s^{2}\right)^{\frac{1}{12}}\;\;\;\;\;,
\label{F_i}
\end{eqnarray}
being $K_{\frac{5}{3}}(\xi)$ a modified Bessel function of the second kind, $\epsilon_{0}$ the vacuum permittivity and $\alpha$ the angle formed by the perpendicular component of the magnetic field with respect to the $e^+/e^-$ momentum. Moreover, $\nu_{c}(\textbf{r},E)$ is the critical frequency 
\begin{eqnarray}
\nu_{c}(\textbf{r},E) =\frac{3e B(\textbf{r})}{4\pi m_{e}} \gamma^{2}(E)\,,
\label{critical_nu}
\end{eqnarray}
around which $e^+/e^-$ emit most of their energy. 
Also, the magnetic field $B(\textbf{r})$ considered throughout this work is spherically symmetric yielding 
\begin{eqnarray}
B(r) = B_{0} \exp(-r/r_{c}),
\label{magnetic_field}
\end{eqnarray}
where $B_{0}$ is the magnetic field strength and $r_{c}= 0.22$ kpc is the core radius of Draco. \\

The electron/positron number density $\psi_{e^-/e^+}(\textbf{r},E)$ produced by DM annihilations and responsible for emitting such synchrotron radiation, is obtained from Eq. (\ref{Sol_dif}).
Now, we can define the total emissivity
\cite{Colafrancesco:2006he,McDaniel:2017ppt,Natarajan:2015hma}
\begin{eqnarray}
j(\nu,\textbf{r})&=&\int^{M}_{E} {\rm d}E \left(\psi_{e^+} + \psi_{e^-}\right)P_\text{syn}(\nu,\textbf{r},E)\nonumber\\
&=&2 \int^{M}_{E} {\rm d}E\,\psi(\textbf{r},E) P_{\text{syn}}(\nu,\textbf{r},E),
\label{j_nu}
\end{eqnarray}
that takes into account the contribution of all $e^+/e^-$ at different energies for a given frequency of emission. 

\begin{figure*}
\centering
    \includegraphics[width=0.6\textwidth]{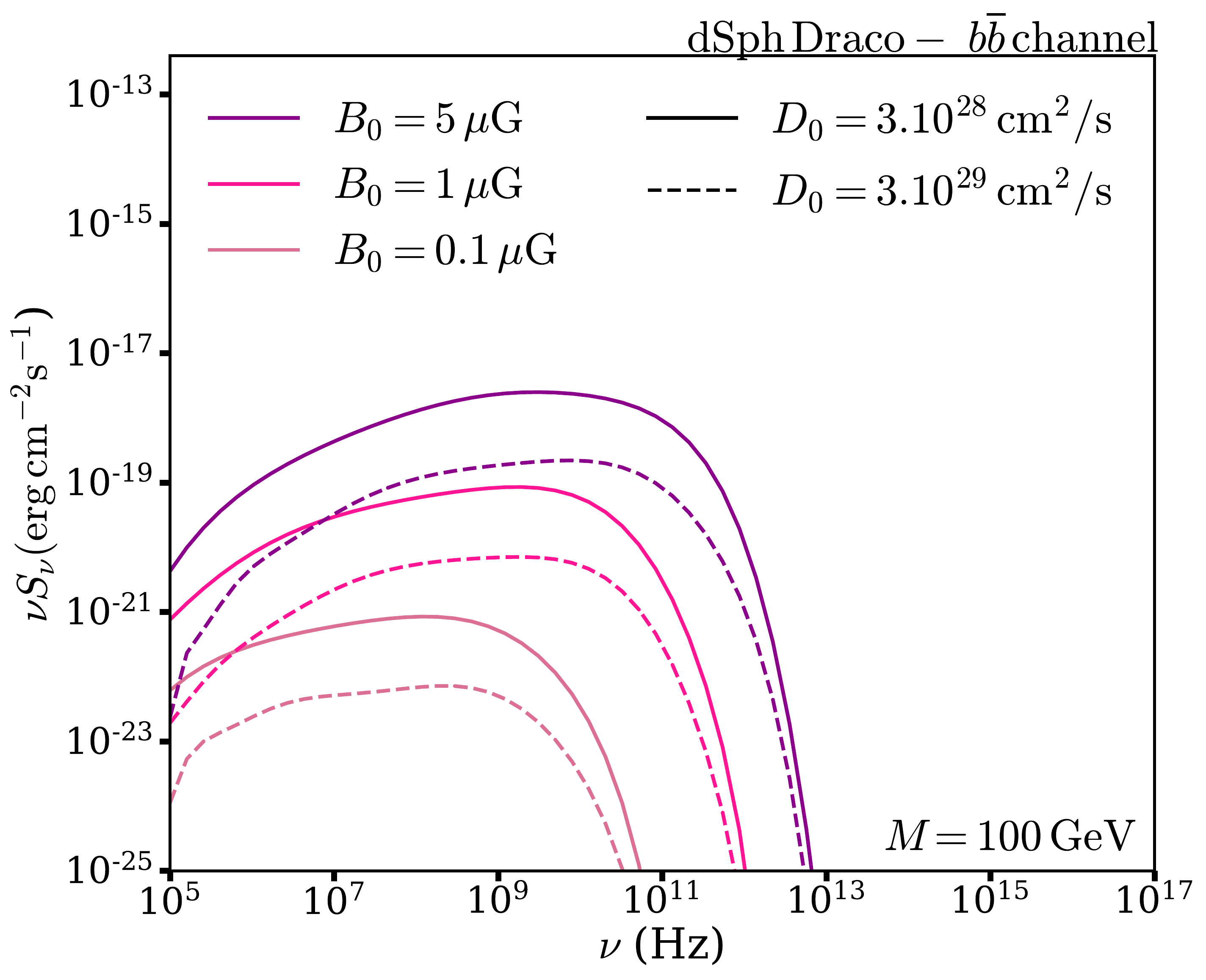}
\caption{Flux density $S_{\nu}$ as given by Eq. (\ref{S_syn}) times frequency $\nu$ for $\Omega_{\text{DRACO}}$ in the range of frequencies $10^5-10^{17}$ Hz. This quantity turns out to be highly dependent of the chosen magnetic field.
Although in the bulk of the article only a moderate magnetic field strength $B_{0}=1$ $\mu$G will be considered, two other values of $B_0$ have been depicted here for illustrative purposes.  The effect of the diffusion has been also depicted, showing that for smaller values of $D_{0}$ the signal increases. The annihilation channel is $b\overline{b}$ for DM with mass $M=100$ GeV and a thermally averaged cross section equals to $ 3\cdot10^{-26} \text{cm}^{3}/\text{s}$.}
\label{Magnet}
\end{figure*}
Thus, the specific intensity $I_{\nu, \theta}$ can be determined just by integrating $j(\nu, \textbf{r})$ along the line of sight (l.o.s.), namely
\begin{eqnarray}
 I(\nu, \theta)\,=\,\int_{{\rm l.o.s.}} {\rm d}l\, \frac{ j(\nu, l, \theta)}{4 \pi},
\label{I_syn}
\end{eqnarray}
where the limits of integration are $l_{\text{max}/\text{min}}=d\cos\theta\pm\sqrt{r_{h}^2-d^2\sin^2\theta}$. 
 Furthermore, the flux density over the solid angle $\Omega$ is given by
\begin{eqnarray}
 S(\nu)=\int_{\Omega} {\rm d} \Omega \, I(\nu, \theta)\,,
 %
\label{S_syn}
\end{eqnarray}
where $\Omega=2\pi(1-\cos\theta)$. In Figure \ref{Magnet} we show how the radio emission strongly depends on the magnetic field. In the case of Draco, the magnetic field lies around $\sim 1$ $\mu$G \cite{Chyzy:2011Ch}. The flux density $ S_{\nu}$ can be computed by integrating over the selected beam size (for more details on the SKA telescope and detection strategies, see Sec. \ref{sec:SKA}). In fact, in radio astronomy the wavelengths  (from mm to a thousand km) are usually comparable to the size of both antennas and baselines, allowing us to adopt interferometric strategies. Different antenna configurations, and hence different baselines, are instrumental to reconstruct the desired image of the sky.
Radio telescopes can map the specific intensity $I_{\nu}(\theta)$ by resorting the so-called van Cittert-Zernike theorem \cite{thompson2017van}. Such a result relates to the cross-correlation map of detected signals between antenna pairs in the projected baseline plane,
and the specific intensity $I_{\nu}(\theta)$  in the sky coordinates through the aperture synthesis technique \cite{Wilner2015sa, Wilner2015se}.
 Images with high resolution are also crucial in the $ S_{\nu}$ determination in order to neither overestimate nor underestimate the signal and to not loose information of the source 
when integrating along the beam coordinates. However, high resolutions would imply a decrease of the sensitivity, crucial to detect faint radio sources as dSphs. 

In the upcoming sections we shall consider the following two solid angle scales:\\

$\bullet\,\,\,\,\,$The solid angle subtended by Draco dSph, dubbed $\Omega_{\text{DRACO}}$, ensuring that all the emission from Draco occurs within this solid angle. In our study we took $\theta_\text{DRACO}= 1.2$ deg.\\

$\bullet\,\,\,\,\,$ The largest angular scale of the radio interferometer $\Omega_{\text{SKA}}(\lambda)$ that is determined by the minimal baseline through the expression $\theta_{\text{max}}(\lambda)=58.61 \lambda/D_{\text{min}}$ (in deg) in which we take $D_{\text{min}}=30$  m for SKA1. Indeed, $\Omega_{\text{SKA}}(\lambda)$ would represent more realistic scenarios than $\Omega_{\text{DRACO}}$, since the former includes the SKA limitations at different frequencies given by $\theta_{\text{max}}(\lambda)$. In this regard, an illustrative value for $\theta_{\text{max}}$ at 150 MHz lies on 3.90 deg, while for $\nu=14$ GHz $\theta_{\text{max}}$ becomes 0.042 deg. This fact indicates that the response of SKA with the frequency would affect the received flux density $S_{\nu}$ according to the integration limits for $\Omega_{\text{SKA}}(\lambda)$ in Eq. (\ref{S_syn}).


\section{SKA telescope}
\label{sec:SKA}

In this Section, we will provide some technicalities on the SKA telescope and detection strategies. 

%
In particular, the SKA1-LOW consists of a low-frequency aperture array located in Australia, while the SKA1-MID is a mid-frequency array of reflector antennas, placed in South Africa. The SKA1-LOW is conformed by approximately $131,000$ log-periodic dual polarised antennas, some of them forming a compact core of $1$ km and the rest grouped in about $512$ stations of $10$s meters of diameter each distributed over $40$-km.  Its range of frequencies, characterised by the lengths of the shortest and longest dipole antenna elements, lies between $50$ MHz and $350$ MHz. Once antennas collect the radio waves, a time delay is introduced with the purpose of reconstructing the signal in a particular direction of detection constituting a signal beam and then, the signals between pairs of antennas are cross-correlated through the correlator. In addition, by combining different sets of timing delays it is possible to construct independent signal beams covering a large FoV and increasing the survey speed. 
Once the signal is processed, a visibility map can be set and hence the information of the specific intensity of the source obtained. In the case of the SKA1-MID the frequencies of detection are divided into five bands ranging from $350$ MHz to $\sim 13.8$ GHz, though in principle the dishes are prepared to measure frequencies until $20$ GHz. In addition, SKA1-MID includes $64$ $13.5$-m diameter dishes from the MeerKAT (SKA1 precursor) and $190$ $15$-m built specifically for the SKA1. Its antennas will be distributed in a compact core with a diameter of ~1 km, a further 2-dimensional array of randomly placed dishes out to $\sim 3$ km radius and three spiral arms until a radius of ~80 km. After measure, signals need to be cross-correlated with each other and processed for further analysis. \\

Both high resolution and sensitivity are critical in the detection of radio faint sources, being SKA1 able to attain both goals.
 Concerning the former, the spatial configuration of the SKA1 allows us to develop interferometry strategies to reach a high angular resolution. Even though radio astronomy searches present more limitations  in this matter than searches performed in other frequencies, the SKA1 long baselines ensure one of the highest angular resolutions 
given by the Rayleigh criterion,  $\theta_{\text{res}}(\lambda)= 58.61 \lambda/D_{\text{max}}$  (in deg) where $\lambda$ is the wavelength of the incoming radiation and $D_\text{max}$ is the maximum baseline, for a specific array configuration and Point Spread Function (PSF). At $350$ MHz, in the limit between SKA1-LOW and SKA1-MID, the angular resolution $\theta_{\text{res}}$ $\sim 0.001$ deg for the SKA1-LOW, taking into account $D_{\text{max}}=50$ km and  $\theta_{\text{res}}$ $\sim 2.5 \cdot 10^{-4}$ deg for the SKA1-MID with a baseline of $D_{\text{max}}=200$ km. 
Even though larger baselines mean higher resolutions, the sensitivity could be compromised unless both the spatial configuration and the effective area $A_{e}$ of the interferometer are optimised with respect to the frequency of observation.  In fact, given a bandwidth $\Delta\nu$ in an interval of time $\tau$, one can determine the sensitivity $S_{\text{min}}$, i.e. - roughly speaking - the minimal detectable flux  of a signal, via the so-called radiometer equation: 
\begin{eqnarray}
 S_{\text{min}}=\frac{2 k_{b} T_{\text{sys}}}{\eta_{s}A_{e}(\eta_{\text{pol}}\tau\Delta\nu)^{1/2}}\,,
\label{S_min}
\end{eqnarray}
where $k_{b}$ is the Boltzmann constant, $\eta_{\text{pol}}$ is the number of polarisation states, and $\eta_{s} = 0.9$ the efficiency of the system.  In principle, as mentioned in \cite{Dewdney:2013ard}, SKA detection 
would be able to analyse integration times of order $1000$ hours and hence, this would be the longest integration time $\tau$ we use to determine the minimal detectable flux along a bandwidth $\Delta\nu$ of $300$ MHz. \\ Concerning  the system noise temperature, 
\begin{eqnarray}
T_{\text{sys}}=T_{\text{sky}}+T_{\text{rcvr}}\,,
\label{Temperature}
\end{eqnarray} 

\begin{figure*}
\centering
   \includegraphics[width=0.496\textwidth]{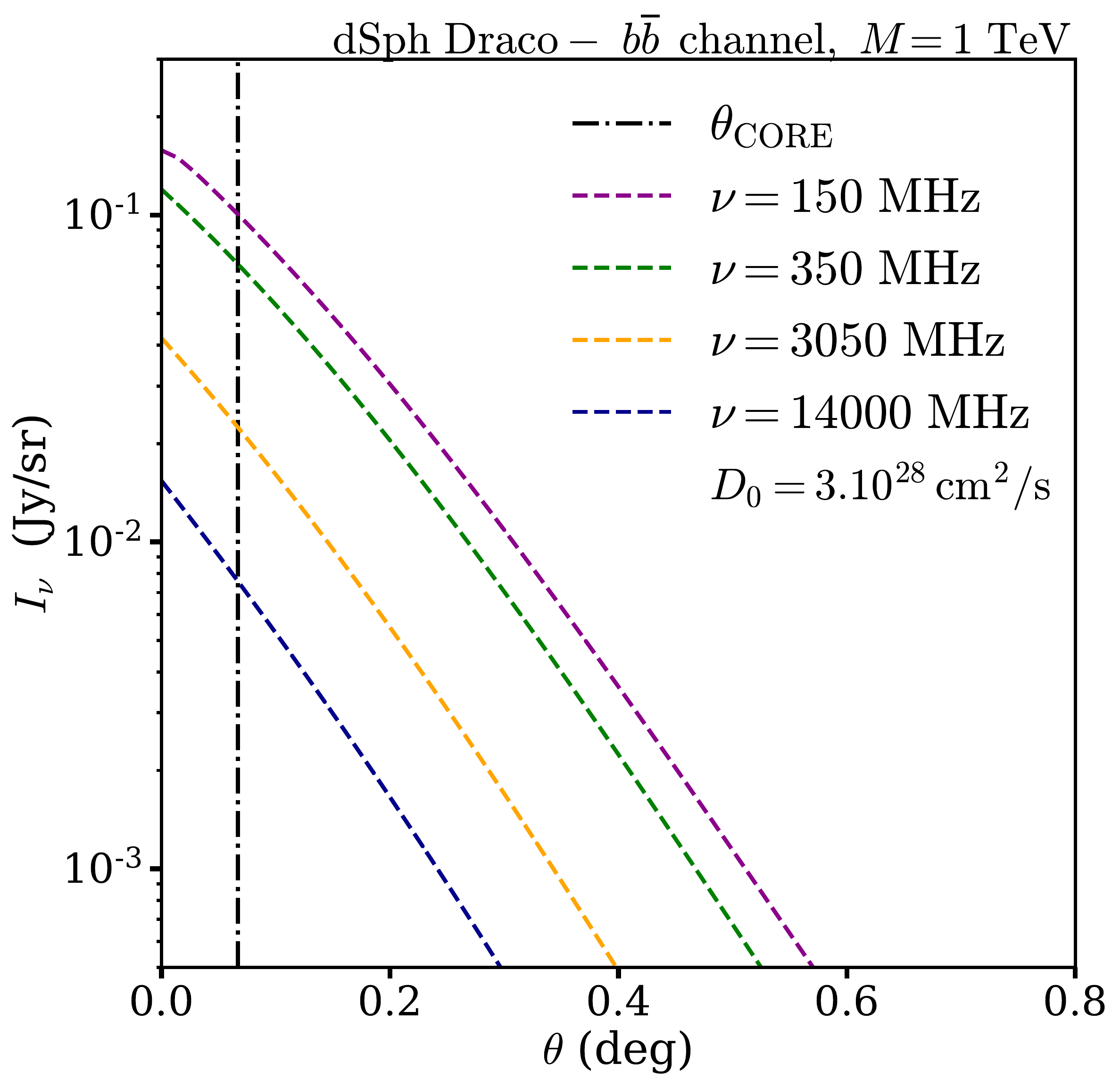}
   \includegraphics[width=0.496\textwidth]{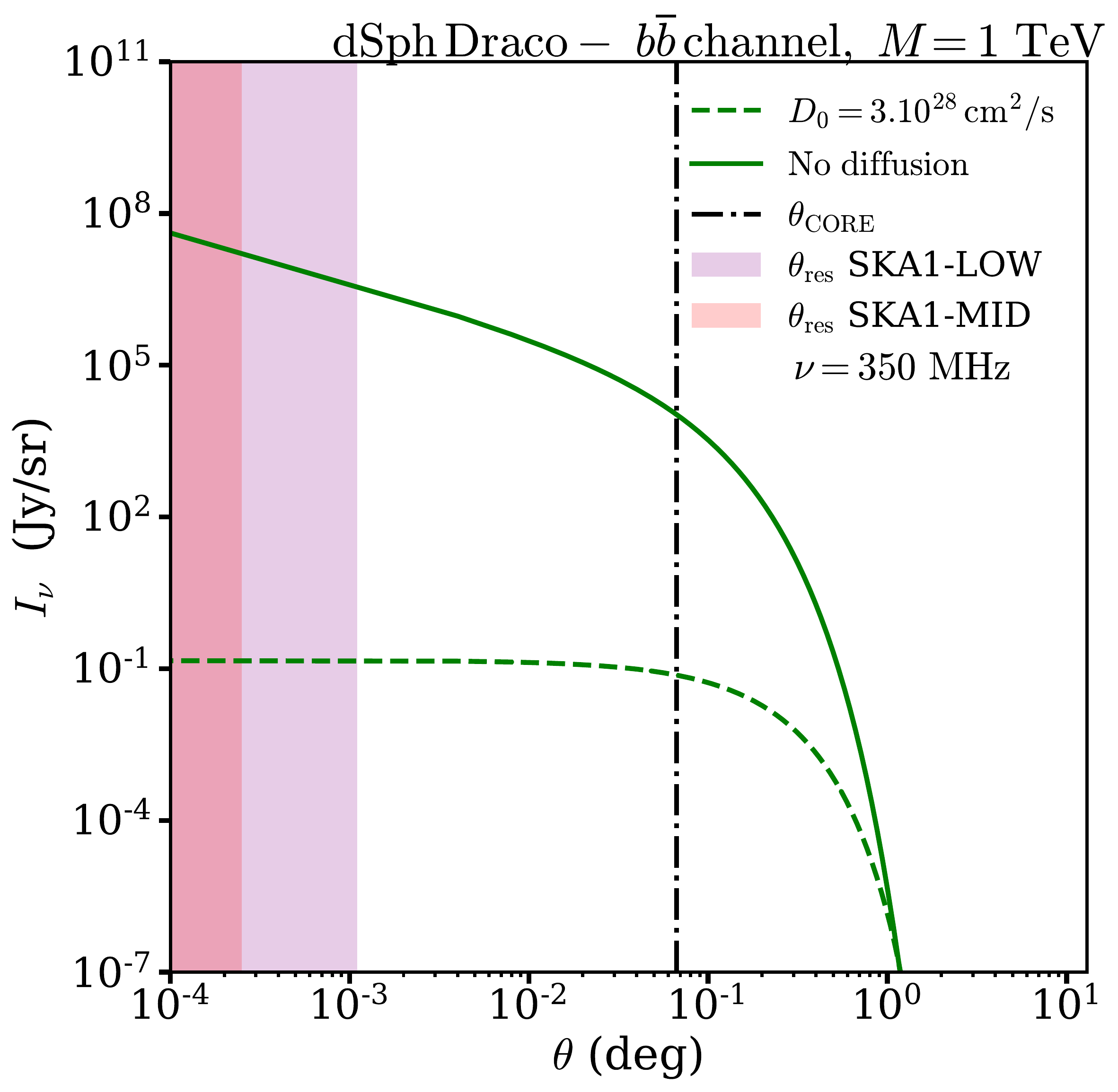}
\caption{{\it Left panel:} specific intensity of the signal $I_{\nu}(\theta)$ at different frequencies for a $M=1$ TeV DM model annihilating into $b\overline{b}$ channel. The vertical dash-dotted black line, represents the core radius of Draco. Obviously the emission exceeds this limit.  {\it Right panel:} the same specific intensity ($\nu=350$ MHz, $M=1$ TeV) for $b\overline{b}$ channel for small values of $\theta$. 
Purple and red bands show the resolution,  $\theta_{\text{res}}$, of SKA1-LOW and SKA1-MID respectively for that frequency. 
 } 
\label{Brightness}
\end{figure*}
it includes the contribution of the sky noise $T_{\text{sky}}$, i.e., all the emission from the sky which does not correspond to the target we want to measure - in the case of a specific observation, or by using the sky noise defined by a uniform model $T_{\text{sky}}=60\lambda^{2.55}$ assumed in the baseline design document \cite{Dewdney:2013ard}, and the receiver noise, also dubbed instrument noise, $T_{\text{rcvr}}$. The latter only dominates for SKA1 in frequencies smaller than $250$ MHz and it is taken as a $10\%$ of $T_{\text{sky}}$ plus a constant temperature of $40$ K. Finally, the effective area $A_{e}$ depends on the frequency and the gain of the receptor. In the case of SKA1-LOW, two different behaviours have to be considered for frequencies lower and higher than $110$ MHz, depending on the sparse-dense transition of the detector. The sparse transition is produced when antenna elements are more distant than around one wavelength
and they act independently making the $A_{e}$ proportional to $\lambda^{2}$. In the dense transition (separation smaller than about $0.5$ times the wavelength) the interaction between the antenna elements force $A_{e}$ to be constant.
Once again, we refer to \cite{Dewdney:2013ard} for further details.
%

%

%
%

\section{Sensitivity Analysis}
\label{sec:Sensitivity}

In this Section, we study the sensitivity of SKA1 to the synchrotron radio emission expected by 
thermal TeV DM annihilation events in the Draco dSph. More specifically, we  compute the specific intensity $I_{\nu}(\theta)$ and the flux density $S_{\nu}$ for the Draco dSph galaxy. In order to plot the sensitivity curves for 10, 100 and 1000 hours we tabulate the System Equivalent Flux Density SEFD=$2k_{b}T_{\text{sys}}/A_{e}$  from \cite{Braun:2017hi} so Eq. (\ref{S_min}) can be used to determine the minimal detectable flux,  $S_{\text{min}}$.

\begin{figure*}[htbp] 
    \centering
       \includegraphics[width=0.496\textwidth]{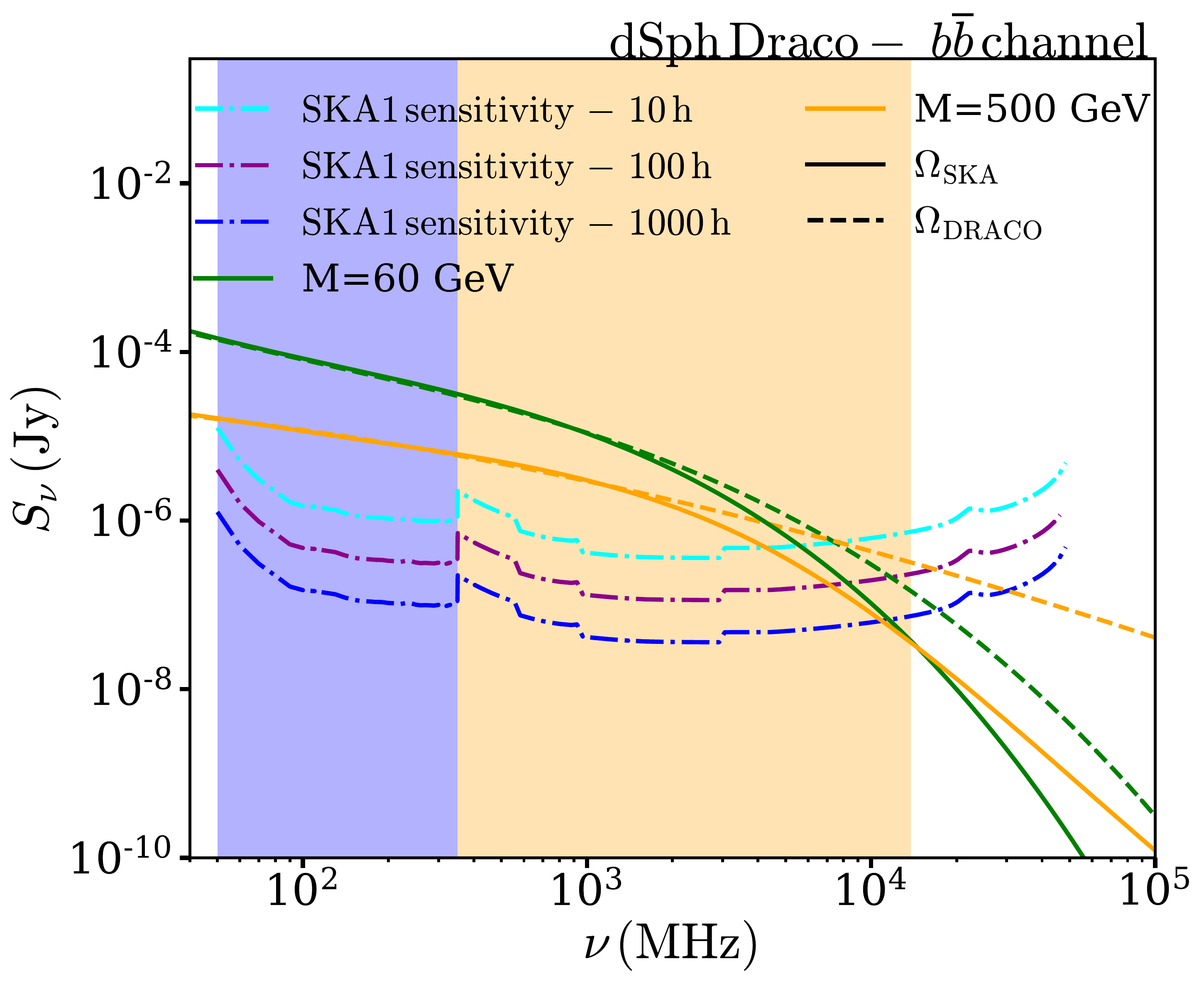}
        \includegraphics[width=0.496\textwidth]{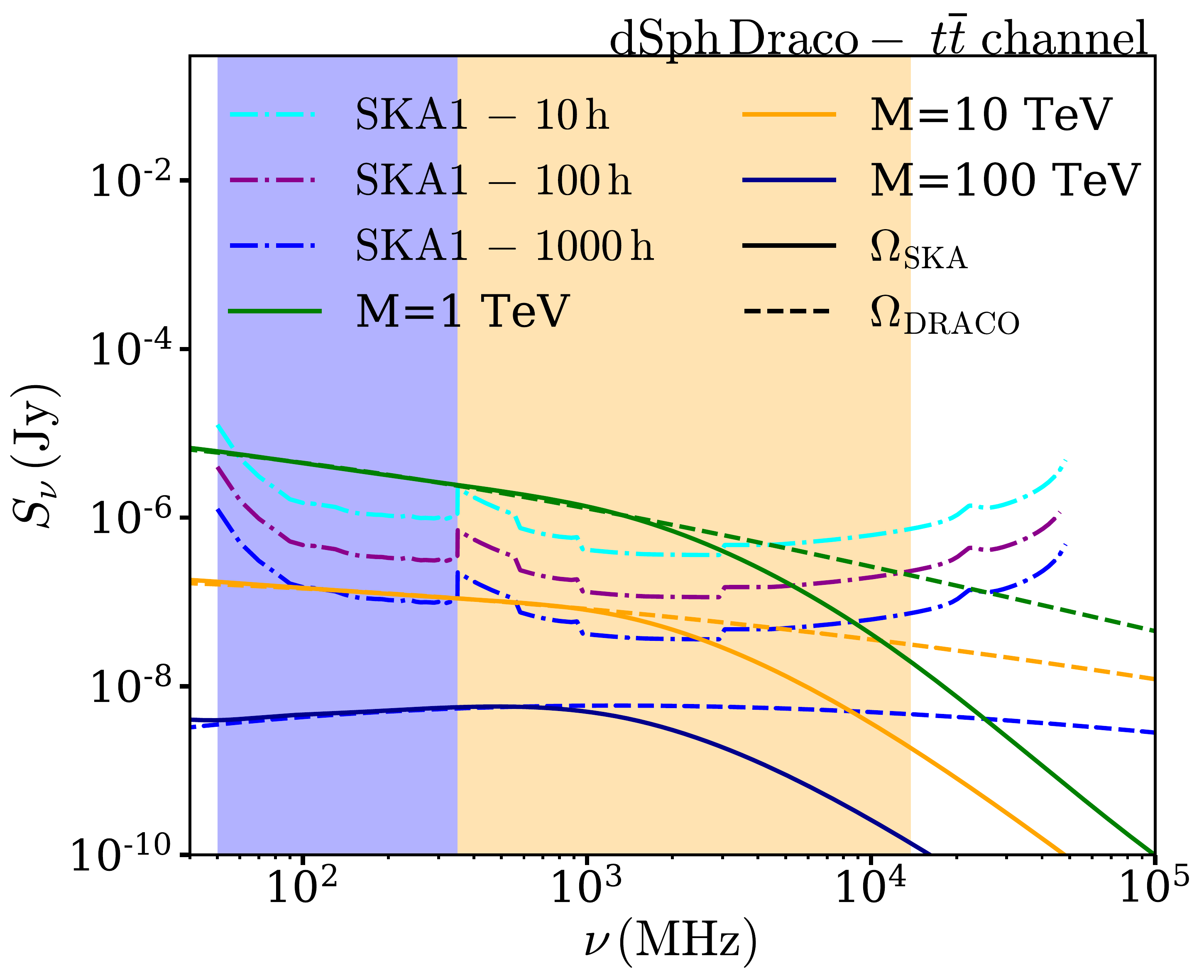}
        \includegraphics[width=0.496\textwidth]{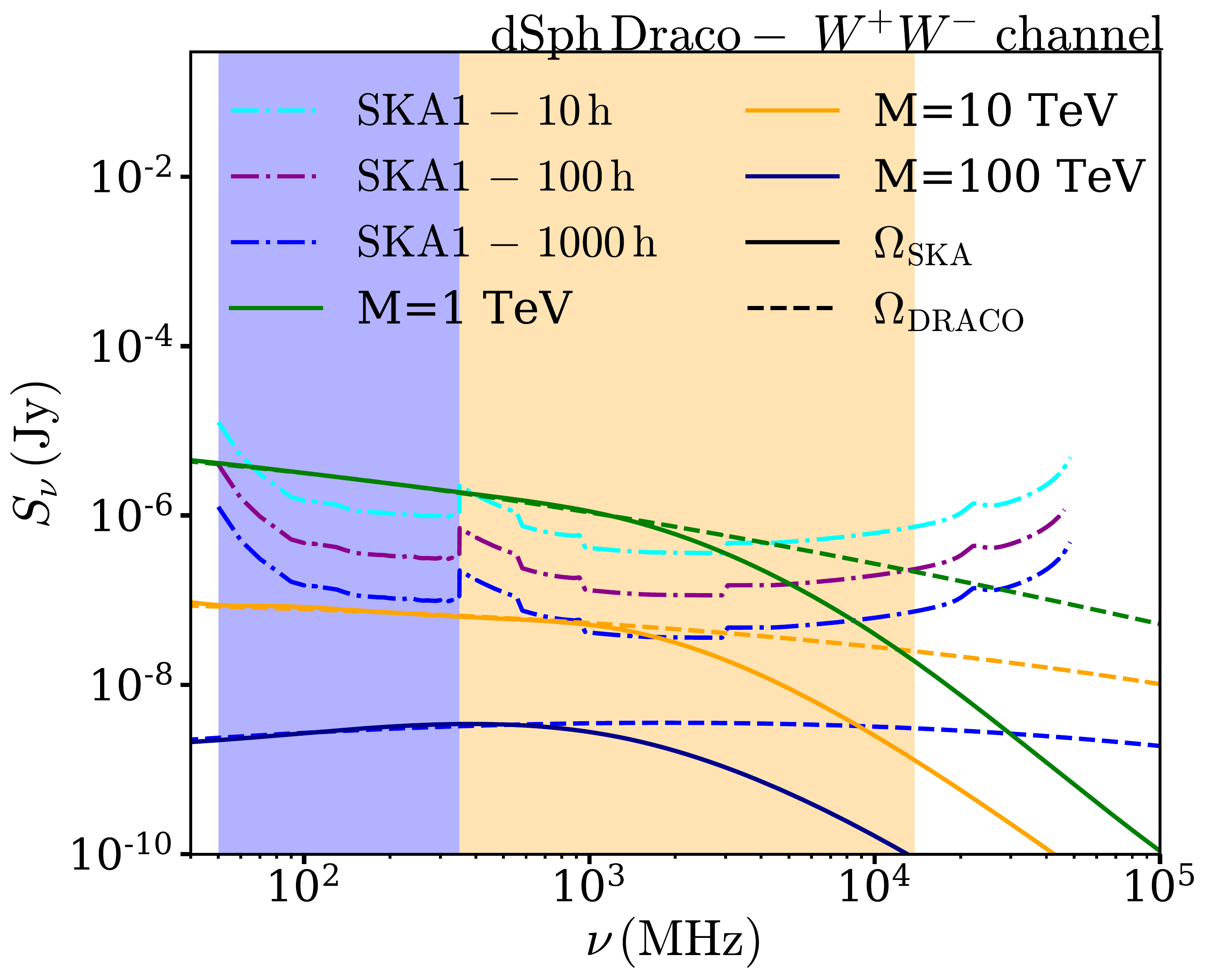}
        \includegraphics[width=0.496\textwidth]{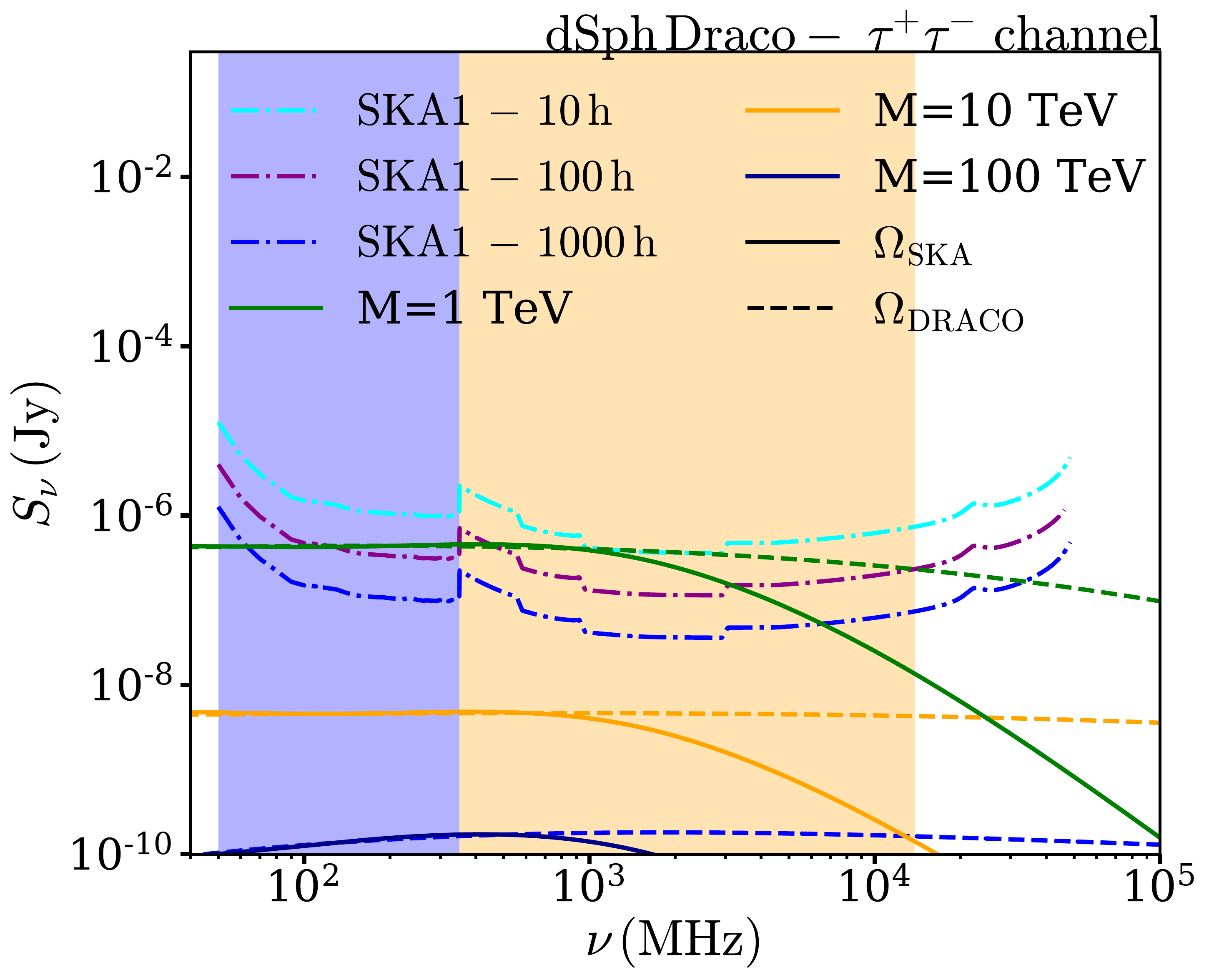}
\caption{Flux density radio emission $S_{\nu}$ for several annihilation channels with 
the canonical thermally averaged cross section of $3\cdot10^{-26} \text{cm}^{3}/\text{s}$ 
and $\beta_j=1$. Color bands correspond to the energy (frequency) range covered by SKA1-LOW (blue) and SKA1-MID (orange), respectively. Dash-dotted lines represent SKA1 sensitivity for different integration times $\tau$,  10, 100 and 1000 hours, and
have been obtained following \cite{Braun:2017hi}. Solid lines represent the flux density of radiation integrated over a solid angle $\Omega_{\text{SKA}}(\lambda)$ while dashed lines correspond to the solid angle $\Omega_{\text{DRACO}}$.
\textit{Upper-left panel: } Sub-TeV DM annihilating into $b\overline{b}$ channel with masses of 60 GeV and 500 GeV according to \cite{Colafrancesco:2015ola}. 
\textit{Upper-right and lower-left/right panels:} Flux density of radiation for DM annihilating into $t\overline{t}$, $W^+W^-$ and $\tau^{+}\tau^{-}$ channels. The first two scenarios would be detectable for masses below 10 TeV, while even smaller DM masses would be required
to be detected through 
the $\tau^{+}\tau^{-}$ channel.
Also, for these three panels lighter DM candidates appear to be  better detected with SKA1-LOW, given the fact the flux density is higher at low frequencies. For heavier DM masses, the flux density turns into an almost-flat straight line provided the integration is performed over $\Omega_{\text{DRACO}}$.
} 
\label{Synchro_SM}
\end{figure*}

\subsection{Model-independent thermal TeV DM}
\label{subsec:Model-independent}

Herein we consider model-independent TeV DM candidates annihilating $100\%$ in a single SM channel, i.e., $\beta_j=1$ in Eq. (\ref{Q}), with the benchmark thermally averaged annihilation cross section $\langle \sigma v \rangle = 3\cdot10^{-26} \text{cm}^{3}/\text{s}$. \\

First of all, let us shed some light on key aspects of specific intensity $I_{\nu}(\theta)$. 
 In Figure \ref{Brightness}, we plot the specific intensity $I_{\nu} (\theta)$ against the angle $\theta$. 
As mentioned in Section \ref{sec:Sync}, the diffusion model is spherically symmetric. In both panels, we present the specific intensity for 1 TeV DM particles annihilating into the $b \overline{b}$ channel. 
The left panel enables a straightforward comparison of the diffused emission at different frequencies that either SKA1-LOW (150 and 350 MHz) or SKA1-MID (350, 3050 and 14000 MHz) will cover. For high frequencies the angular extension of the emission is confined in a small region while the signal at lower frequencies reaches larger angles. Considering this fact, even though the highest intensity lies inside the core of Draco, estimated as $\sim 0.067$ deg \cite{fomalont1979limits}, a large proportion of the signal extends further in radius. Nonetheless, the signal follows the expected peak at the centre given by the DM density profile smoothed by the diffusion. In the right panel we also show the specific intensity at $350$ MHz (which establishes the limit between SKA1-LOW and SKA1-MID) 
with and without spatial diffusion. Therein, it is possible to observe that neglecting diffusion results in a signal concentrated at the centre of Draco, since in this case the $e^{+} /e^{-}$ number density follows the DM density profile strictly. On the other hand, it is also worth noting that in either case (with and without diffusion) the profile of the radio signal would have a spatial dependence if we consider a spatially-dependent magnetic field. Here, this effect is assumed to be negligible due to the small radii considered in this work.
%
%
The right panel of Figure \ref{Brightness} shows the angular resolution of SKA1 with respect to the extension of the source. 
Once the diffusion is considered, the radio emission extends further in distances, decreases slightly with the angle $\theta$. 
In fact, we have seen - in the left panel of Figure \ref{Brightness} - that 
%
whenever diffusion is considered, most of the signal does not come from the centre and hence the source cannot be thought of as being point-like. In this regard, we conclude that both SKA1-LOW and SKA1-MID can resolve part of the structure of the emission for Draco assuming an angular resolution $\theta_{\text{res}}$ $\sim 0.001$ deg for SKA1-LOW and $\sim 2.5 \cdot 10^{-4}$ deg for SKA1-MID at $350$ MHz.


%
The specific intensity $I_{\nu}(\theta)$ and the high resolution angle typical of radio telescope, gives information about the spatial distribution of the radio emission detected in the target - if any. In the case of no detection of any signal, 
we are more interested in the sensitivity of the instrument and the flux density $S_{\nu}$: by integrating $I_{\nu}(\theta)$ over the solid angle $\Omega$ according to Eq. (\ref{S_syn}), we obtain the radio flux density $S_{\nu}$ expected by TeV DM candidates annihilating into a specific channel and 
we can set constraints on the DM mass and annihilation cross section.
%

First, we check the agreement of our results with previous studies and we plot in the upper-left panel of Figure \ref{Synchro_SM} the densities $S_{\nu}$ for two sub-TeV DM masses annihilating into $b\overline{b}$ channel, using 
the benchmark thermally averaged cross section value $\langle \sigma v \rangle = 3\cdot10^{-26} \text{cm}^{3}/\text{s}$. Dashed lines correspond to the flux density integrated over the solid angle subtended by Draco, $\Omega_{\text{DRACO}}$, and solid lines represent the emission integrated over the SKA angle, $\Omega_{\text{SKA}}(\lambda)$, defined in Section \ref{sec:Sync}. In more detail, given a certain baseline $D$, the angle of the radio detector increases with the wavelength, in other words, the integration beam is smaller for higher frequencies. Due to the strategy of this analysis, the loss of the signal at high frequencies is clear in 
 the four panels of Figure \ref{Synchro_SM}.
%
\,At this stage, let us emphasise that the flux density shows a steepness at lower frequencies, implying that SKA1-LOW (blue region) will be more competitive features in order to detect DM candidates of lower masses. As the DM mass increases, (see e.g. the DM candidates with  $M=1$ TeV and $M=10$ TeV in the upper-right panel for the $t\overline{t}$ channel) the slope is less pronounced and SKA1-MID (yellow region) will be more competitive in order to set constraints on the DM candidate, due to the slightly
better sensitivity. Therein, dotted-dashed lines of different colours hold for different exposition times: in particular cyan (10h), magenta
(100h) and blue (1000h). 
%
%
%
%
%
This change in the slope occurs because heavy DM mass enables $e^{+}/e^{-}$ to reach higher energies and thus the frequency of the peak,  which is proportional to the kinetic energy $E$ according to Eq. (\ref{critical_nu}), is shifted to higher frequencies.
Finally, In the lower-left and lower-right panels in Figure \ref{Synchro_SM}, we show the SKA detectability to TeV-DM annihilating into  $W^+W^-$ and $\tau^+\tau^-$ channels, respectively. TeV DM annihilating in the leptonic channel, i.e. $\tau^{+}\tau^{-}$ seems to be the most disfavored scenario for radio detection. 
In fact, according to the $S_{\text{min}}$ limit as given by Eq. \eqref{S_min}, lower-right panel shows that masses of 10 TeV remain inaccessible even if the integration time $\tau$ is increased of one order of magnitude. 
Instead, DM particles annihilating in the $t\overline{t}$ and $W^+W^-$ channels, are more favored to be detected by SKA1 - e.g.
\,radio signal from 1 TeV DM particles could be detected with 10 hours of data-taking. Nonetheless, for DM masses around 10 TeV, 1000 hours would be necessary.
%
%

\begin{figure*}[htbp] 
    \centering
        \includegraphics[width=0.7\textwidth]{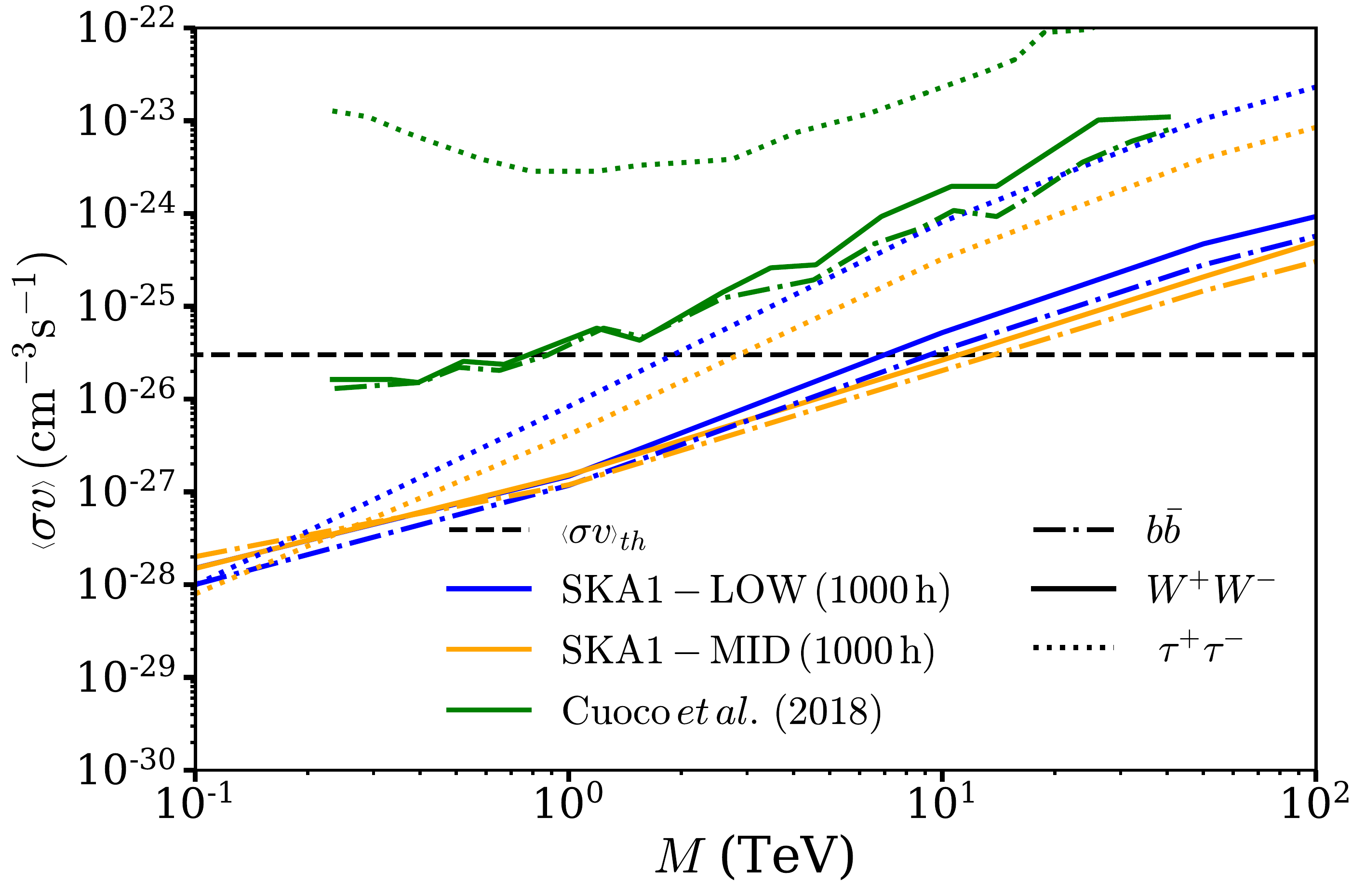}
\caption{Sensitivity constraints on the DM particle cross section for model-independent DM annihilating into $b\overline{b}$, $\tau^{+}\tau^{-}$ and $W^+W^-$ channels with SKA1. 
Horizontal dashed black line represents the cross section limit $\left\langle \sigma v \right\rangle_{th}= 3\cdot10^{-26} \text{ cm}^{3}/\text{s}$ for thermal relics.
The solid angle considered to set this map is $\Omega_{\text{SKA}}(\lambda)$ with a minimal baseline of 30 m.  The lines in blue would be detected within 1$\sigma$ by SKA1-LOW while the lines in orange would be detected by SKA1-MID, after 1000 hours of integration time. For comparison, we plot upper limits from cosmic-ray antiproton data (green lines) \cite{cuoco2018constraining}. For example, the sensitivity constraint for the  $W^+W^-$ channel lies close to $M=7$ TeV for SKA1-LOW and $M=10$ TeV for SKA1-MID. 
} 
\label{Plot_sigmav}							
\end{figure*}

In Figure \ref{Plot_sigmav} we show the constraints on the averaged annihilation cross section $\left\langle \sigma v \right\rangle$ vs the DM mass $M$ for model-independent DM candidates annihilating into $b\overline{b}$ (dotted-dashed line), $\tau^{+}\tau^{-}$ (dotted line) and $W^+W^-$ (full line) channels. As previously stressed, SKA1-MID (orange lines) will be slightly more competitive than SKA1-LOW to constrain DM candidates with masses heavier than 1 TeV. 

\subsection{Black-hole effect in annihilating DM signals}
\label{BH}

Generally speaking, several aspects on both the astrophysical and particle physics side could generate some boost in the secondary fluxes produced by the annihilation of DM particles. In particular, sub-clumps in DM halos \cite{Beck:2015rna}, DM-spike associated to a Black Hole (BH) \cite{Gonzalez-Morales:2014eaa}  and Sommerfeld enhancement \cite{sommerfeld1931beugung,feng2010sommerfeld}  have been widely considered in several studies devoted to DM indirect searches. In fact, DM sub-clumps in galactic satellites could enhance the flux of a factor 
 $2-3$, in both NFW and Burkert DM density distribution profile \cite{Beck:2015rna}. Nonetheless, under the assumption that subhalos and sub-subhalos are tidally stripped 
some authors estimate a smaller multiplicative factor, $\sim 1.3$ (if one assumes $M_{\text{vir}}=7\cdot10^{7}M_{\odot}$ \cite{Beck:2015rna}) times the smooth halo contribution \cite{Moline:2016pbm}.  

In this Section, we will show that the presence of a BH in Draco would result in a radio emission enhancement by several orders of magnitude. 
%
The possibility to have an intermediate-mass BH has been considered in previous work \cite{Colafrancesco:2006he,Gonzalez-Morales:2014eaa}. 
Authors in \cite{wanders2014dark} studied the case where a BH grows adiabatically and thus, the inner part of the DM profile $\rho(r) \sim r^{-\gamma}$ is adiabatically contracted into a final profile $\rho(r) \sim r^{-\gamma_{sp}}$, with $\gamma_{sp}>\gamma$. 
Consequently, 
the modified DM density profile  $\rho_{\text{DM}+\text{BH}}(r)$ in the presence of a BH becomes
\begin{eqnarray}
 \rho_{\text{DM}+\text{BH}}(r) =
  \begin{cases}
              \frac{M}{\langle \sigma v \rangle (t-t_f)} & \text{if $r<r_{cut}$} \\
              \rho_{\text{DM}}(r_{sp})\left(\frac{r}{r_{sp}}\right)^{-\gamma_{sp}}  & \text{if $r_{cut}\leq r < r_{sp}$} \\
              \rho_{\text{DM}}(r) & \text{if $ r \geq r_{sp}$}\,,
  \end{cases}
\label{BHdensity}
\end{eqnarray}
where $\rho_{\text{DM}}(r)$ is the standard DM density profile (NFW for us) without a BH.  
As seen in Eq. (\ref{BHdensity}), the BH effect turns into a spike in the DM profile. 
In the region $r<r_{cut}$, a plateau $\frac{M}{\langle \sigma v \rangle (t-t_f)}$ exists in the DM density distribution due to a more efficient DM annihilation rate. In this expression, $M$ is the mass of DM particle and $(t-t_{f})$ holds the BH age, taken as $10^{10}$ yr for Draco. The transition between the aforementioned upper limit 
and the density in the region where the BH has no influence ($ r \geq r_{sp}$) is gradually modulated by a power law with exponent $\gamma_{sp}=(9-2\gamma)/(4-\gamma)$.
 %
In the following we will consider a BH mass of $10^{3}M_{\odot}$ according to \cite{Reines:2013pia,Moran:2014yea}; BH masses
bigger than $10^7 M_{\odot}$ are dynamically excluded for Draco \cite{Colafrancesco:2006he}. A full discussion on how the presence of a BH in Draco would affect
the multi-wavelength analysis can be found in \cite{Colafrancesco:2006he}.

\begin{figure*}[htbp] 
    \centering
         \includegraphics[width=0.496\textwidth]{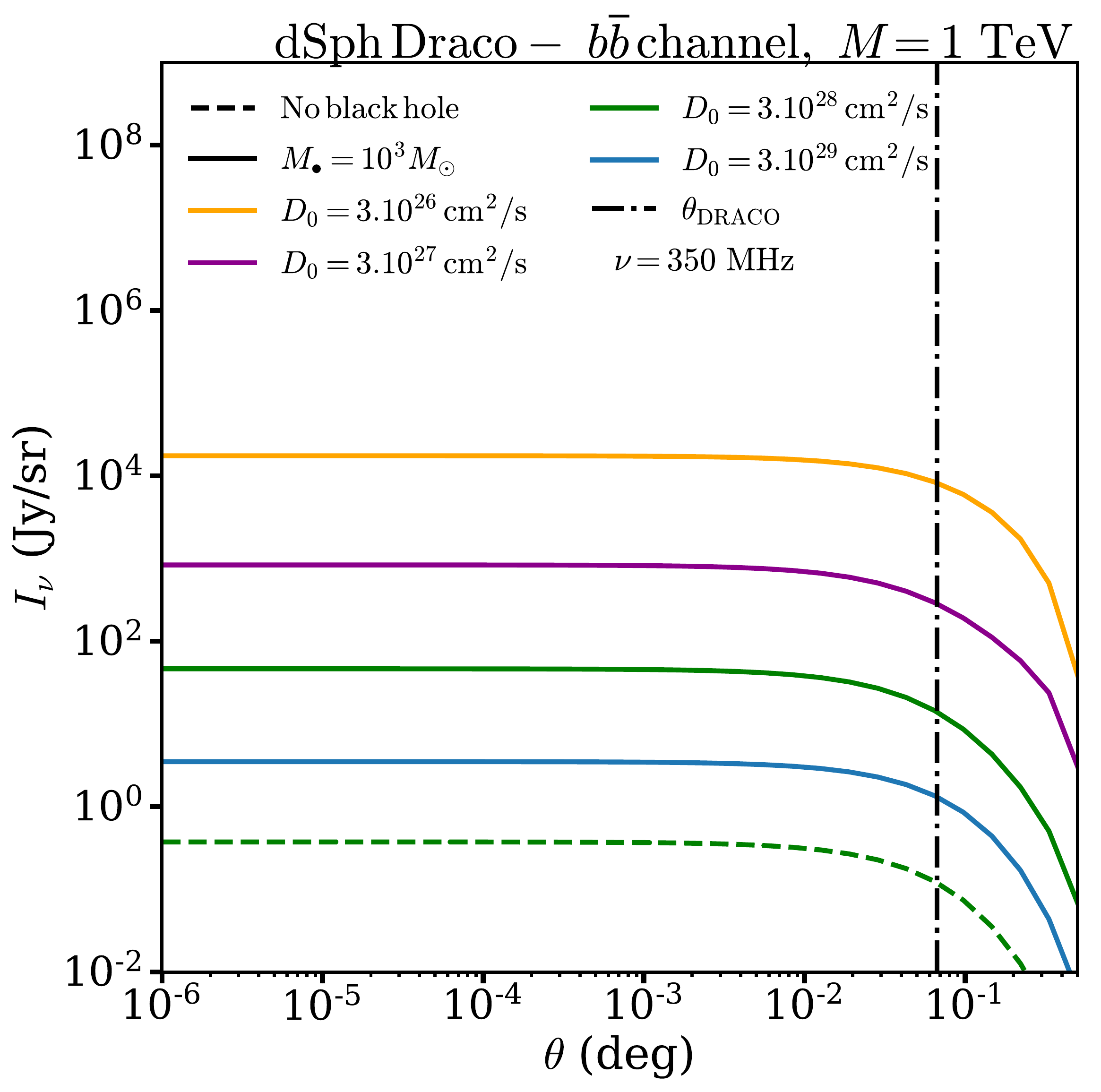}
         \includegraphics[width=0.496\textwidth]{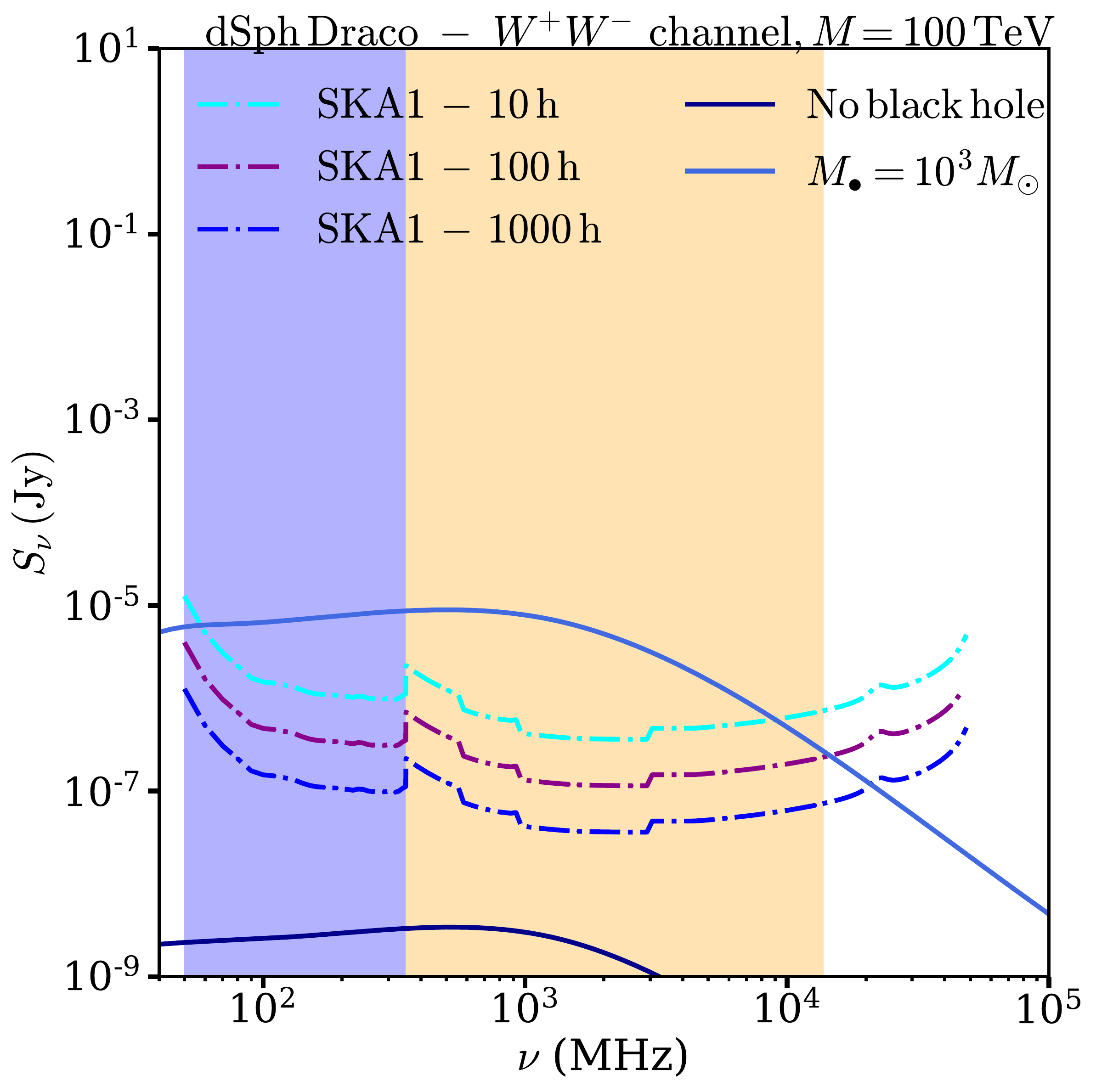}
\caption{ 
 {\it Left panel:} 
 Specific intensity at $\nu=350$ MHz for a $M=1$ TeV DM candidate, both without and with a BH of $M_{\bullet}=10^{3}M_{\odot}$. For the latter case, different diffusion coefficients have been considered. The canonical thermally averaged cross section $3\cdot10^{-26} \text{cm}^{3}/\text{s}$ is considered. 
 As can be seen the emission including a BH exceeds the core of Draco ($\theta_{\rm DRACO} \sim 0.067$ deg).
  {\it Right panel:} 
  Effect in the flux density  when a BH of $M_{\bullet}=10^{3}M_{\odot}$ 
 in the centre of Draco is included. DM mass $M=100$ TeV and annihilation happens 100\% in the  $W^{+}W^{-}$ channel.
As observed there, the signal from TeV DM candidates which, in principle could not be detected by SKA1, would be boosted to SKA1-detectable regions by the inclusion of such a BH, rendering the detection of TeV DM feasible, even for short SKA1 integration times. 
As in  Figure \ref{Synchro_SM}, colour bands correspond to the energy range covered by SKA1-LOW (blue) and SKA1-MID (orange) respectively. 
Dash-dotted lines represent SKA1 sensitivity for different integration times $\tau$ of  10, 100 and 1000 hours.
} 
\label{Synchro_BHplot}							
\end{figure*}

In Figure 6 left panel, we show the specific intensity for a model-independent DM candidate of $M = 1$ TeV that annihilates into $b\bar b$ channel. The enhancement of the signal due to the presence of a BH of $10^{3}M_{\odot}$ in the centre of Draco appears clearly also with strong diffusion, although higher values of the diffusion coefficient would reduce the enhancement in the flux of several orders of magnitude.

For outer regions, the specific intensity for both cases (with and without BH) tends to the same limit. 
%
%
Regarding the flux density, in Figure \ref{Synchro_BHplot} right panel, we study the emission spectra for a model-independent DM candidate with the maximum value of masses considered in this study - i.e.  $M=100$ TeV -  annihilating into $W^+W^-$ channel.  Even considering 1000 hours, SKA1 would not be able to detect such a DM mass. However, for the case of a BH mass $10^{3}M_{\odot}$,  would be detectable with only 10 hours of data-taking. 
We conclude that the presence of a central BH would enhance the synchrotron radio signal significantly without any qualitative differences in the  emission spectra, as showed in the right panel of Figure \ref{Synchro_BHplot}. 

\subsection{Branons and extra-dimensions}
\label{sec:branons}

\begin{figure*}
\centering
    \includegraphics[width=0.496\textwidth]{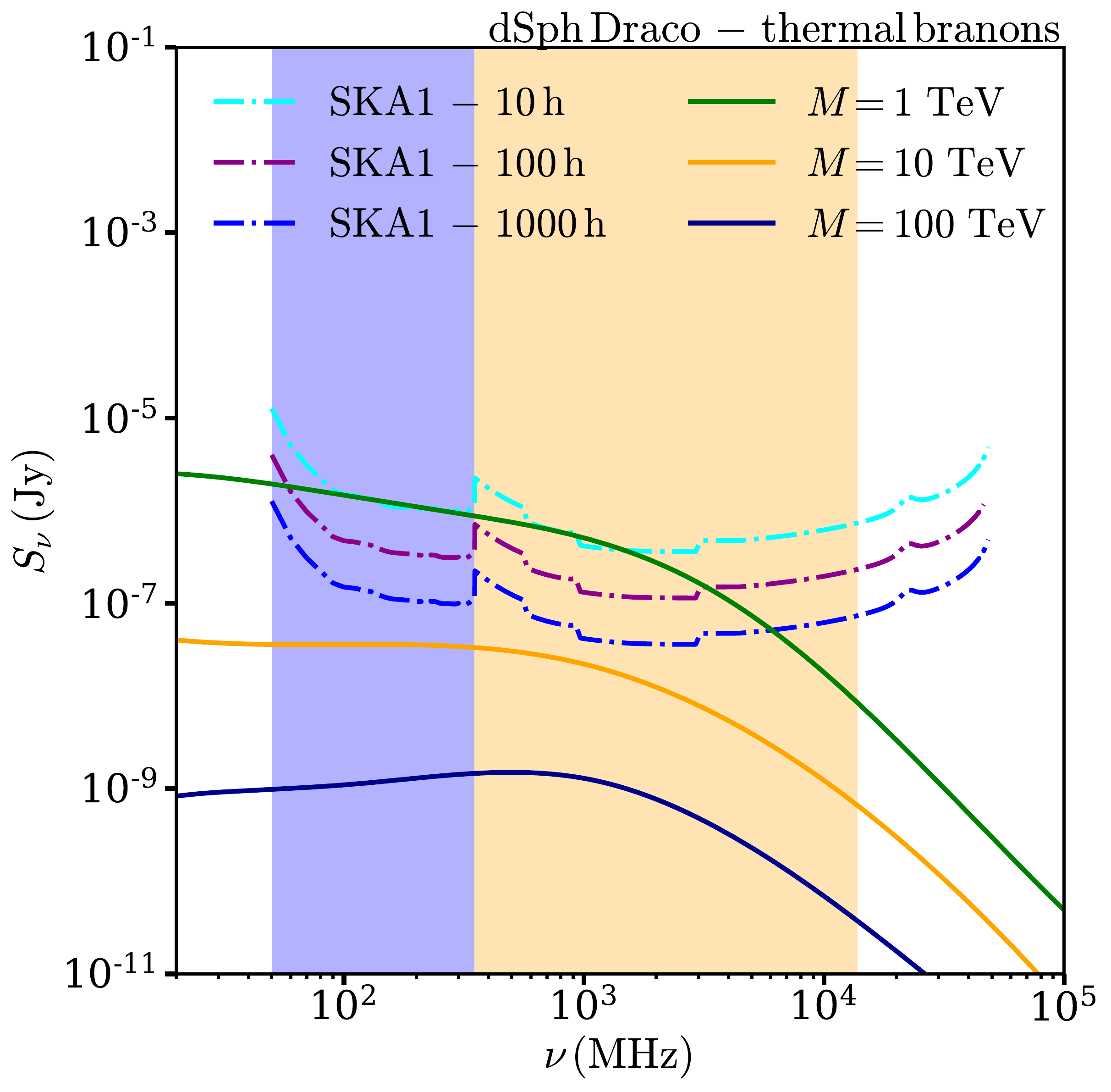}
\includegraphics[width=0.496\textwidth]{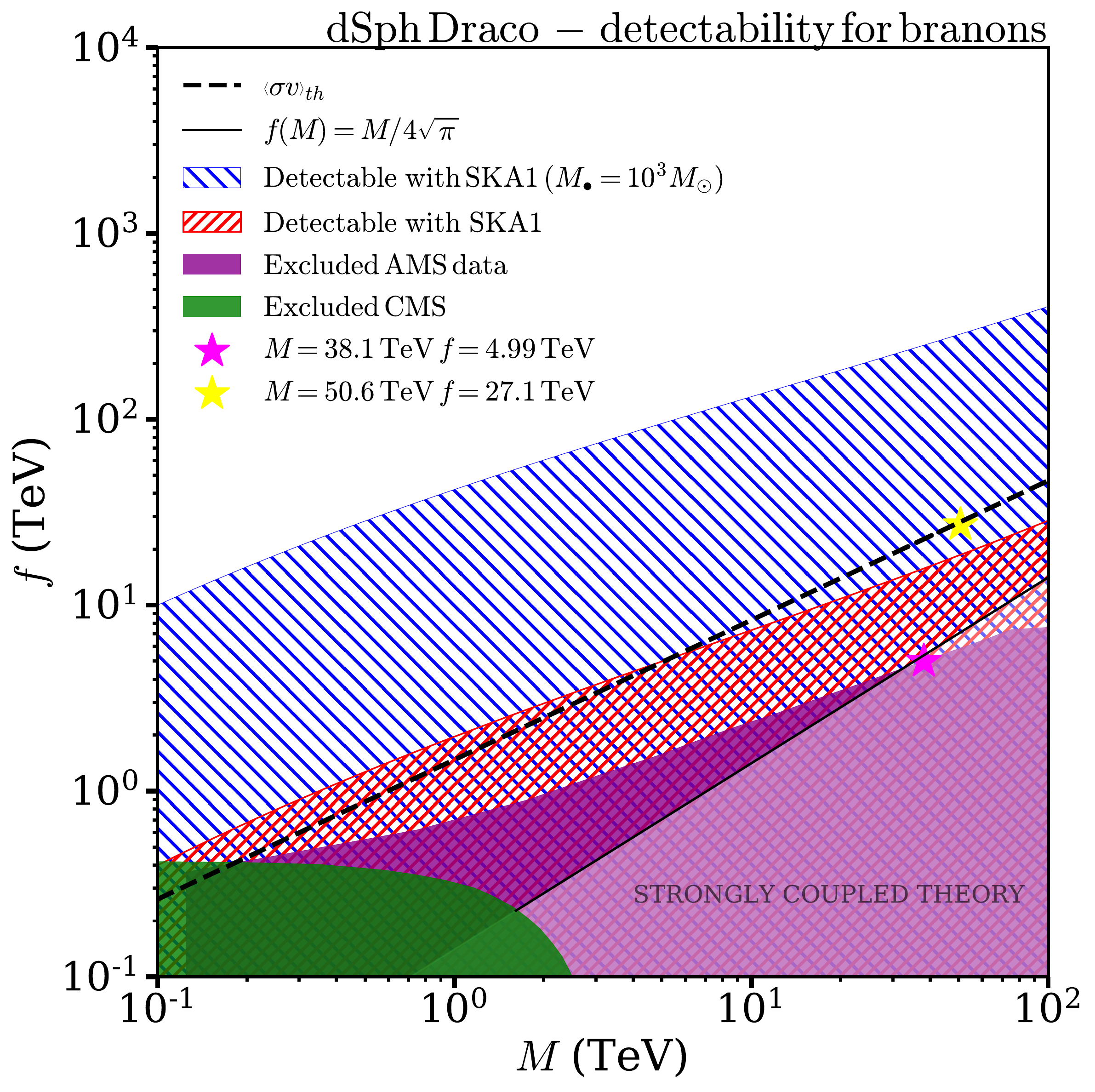}
\caption{
{\it Left panel}: Flux density of radiation $S_{\nu}$  over $\Omega_{\text{SKA}}(\lambda)$ for freeze-out thermal branons \cite{Cembranos:2003mr,Cembranos:2003fu}. For illustrative purposes we considered one extra-dimension ($D=5$).
%
%
%
\,{\it Right panel:}  Detectability of branons from Draco with SKA1 on the $(M,f)$ parameter space after 1000 hours of integration time. The solid angle $\Omega_{\text{SKA}}(\lambda)$ has been considered to set this map with a minimal baseline of 30 m. The dashed black line represents the curve $f(M)$ for a thermal relic with $\left\langle \sigma v\right\rangle_{\text{th}}=3\cdot10^{-26}\,\text{cm}^{3}/\text{s}$. 
Above such a line $\left\langle \sigma v\right\rangle < \left\langle \sigma v\right\rangle_{\text{th}}$, i.e. branons produced by the benchmark freeze-out mechanism would be excluded. In other words, such a region represents the area in which branons would be overproduced, exceeding the cosmological parameters expected by the cosmological standard model $\Lambda \text{CDM}$.   The region below that line represents the parameter space region where $\left\langle \sigma v\right\rangle > \left\langle \sigma v\right\rangle_{\text{th}}$. 
 The region in red could be excluded by future observation of Draco with SKA1 (this work). In more details, SKA1 could be able to exclude branon masses up to $\sim$6 TeV, while heavier thermal branon with masses up to 100 TeV could not be constrained by this kind of observation. On the other hand, if the Draco dSph hosts a BH of $10^{3}M_{\odot}$,  the most part of the parameter space would be excluded. The region in green represents the constraints obtained by CMS to the branons model  \cite{khachatryan2016search}, that are the tightest constraint obtained by colliders so far.  The purple region is excluded by the analysis of AMS  $e^{+}/e^{-}$ data \cite{Cembranos:2019amc}. The shaded white region below the full black line, shows the strongly coupled region, i.e. the area in which our exclusion limits could change due to the validity of the Effective Field Theory. In fact, for that region of the parameter space higher loop corrections become important to branon phenomenology with respect to the tree-level contribution \cite{Cembranos:2005jc}. Finally, the branon DM candidates $M =50.6$ TeV and tension $f=27.1$ TeV (yellow star), and the candidate $M =38.1$ TeV and tension $f=4.99$ TeV (magenta star) have been depicted. 
}

\label{Synchro_candidates}
\end{figure*}

In this Section, we will consider an extra-dimensional model for TeV DM candidates. Roughly speaking, 
brane-world scenarios address the hierarchy problem in Physics by suggesting that our Universe is a $(3+1)$-dimensional brane, where the SM fields propagate. Such a brane is embedded in a higher extra-dimensional compact space - the bulk - that hosts the propagation of gravity. 
Thus, brane-world theories are defined through the manifold $\mathcal{M}_{D}$=$\mathcal{M}_{1+3}\times B$, being $\mathcal{M}_{1+3}$ the brane four-dimensional space-time embedded in a $D-$dimensional bulk, both of them characterised by their respective metric tensors, $\widetilde{g}_{\mu\nu}$ and $g'_{mn}$.
The global space-time of such a manifold can be represented by
\begin{eqnarray}
 {\rm d}s^{2}=\widetilde{g}_{\mu\nu}\left(x\right)W\left(y\right){\rm d}x^{\mu}{\rm d}x^{\nu}-g'_{mn}\left(y\right){\rm d}y^{m}{\rm d}y^{n},
\label{brane_interval}
\end{eqnarray}
where $x^{\nu (\mu)}$ and $y^{m(n)}$, with $\nu, \mu=0,1,2,3$ and $m,n=4,5,6...,D-1$, denote the coordinates for the four-dimensional brane and the compact extra-dimensional space respectively. Here, $W(y)$ is the warp factor along the extra-dimensions. In the case where $W(y)\neq1$, 
a curvature is produced that explicitly breaks the isometry of the brane - that is considered as gauge  symmetry - and the brane fluctuation can be parametrised by a massive (M) pseudo-Goldstone field, $\pi^{\alpha}$, dubbed branon. In the range of low tension $(f\gtrsim M_{})$ 
the branon dynamic is given by the Nambu-Goto action added to the usual SM action, associated with the following Lagrangian \cite{Sundrum:1998sj,Bando:1999di,Dobado:2000gr,Cembranos:2001rp,Alcaraz:2002iu,Cembranos:2004jp};
\begin{eqnarray}
{\mathcal L}_{Br}\,&=&\,
\frac{1}{2}g^{\mu\nu}\partial_{\mu}\pi^\alpha
\partial_{\nu}\pi^\alpha-\frac{1}{2}M^2\pi^\alpha\pi^\alpha
\nonumber\\
&&+\,\frac{1}{8f^4}(4\partial_{\mu}\pi^\alpha
\partial_{\nu}\pi^\alpha-M^2\pi^\alpha\pi^\alpha g_{\mu\nu})
T^{\mu\nu}_{\rm SM}.
\,\label{lag}
\label{brane_action}
\end{eqnarray}
As Eq. (\ref{lag}) shows, the coupling between branons and SM particles is highly suppressed by a factor $f^{-4}$. Also, the brane Lagrangian \eqref{lag}  conserves parity and terms with odd number of branons are not allowed, hence, branons are stable. Being branons weakly interacting, massive and stable, they are natural candidates for DM \cite{Cembranos:2003mr,Kugo:1999mf,Cembranos:2004eb,Maroto:2003gm,Maroto:2004qb}. 
Moreover, branons have the possibility of annihilating into SM particles, whose probability of annihilation is expressed by the branching ratio 
and depends on the branon mass $M$ and the tension of the brane $f$ \cite{cembranos2016disformal}. 
The branons parameter space $(f,M)$ has been studied through different methods in previous works.  For instance, collider detections by the ILC, LHC or CLIC have been taken into account \cite{Alcaraz:2002iu,Cembranos:2004jp,Achard:2004ds,Creminelli:2000gh,Cembranos:2005jc,khachatryan2016search}. Astrophysical and cosmological bounds for brane-world theories were obtained in \cite{Cembranos:2003mr,Kugo:1999mf,Cembranos:2004eb,Maroto:2003gm,Maroto:2004qb}. Other constraints are obtained in previous studies for gamma-ray detectors, e.g. EGRET, FERMI and MAGIC \cite{Cembranos:2013pwa,Cembranos:2011hi} and by studying the positrons flux recently detected by AMS-$02$ \cite{Cembranos:2019amc}.  
Here, we focus on the annihilation of TeV branons producing $e^+/e^-$  as a final state.  
%
%
For this range of mass, $e^+/e^-$ are  mainly produced via a combination of $W^+W^-$ and $ZZ$ gauge bosons and $t\overline{t}$ quark channel, due to the branon branching ratios \cite{Cembranos:2012xp}. 
In the left panel of Figure \ref{Synchro_candidates} we show the synchrotron radio emission produced by the annihilation of thermal branons, i.e. with ${\langle \sigma v\rangle}=3\cdot10^{-26} \text{cm}^{3}/\text{s}$. Without any boost, the SKA1 detection of thermal branons for 1000 hours of integration time becomes feasible for masses smaller than 6 TeV.
%
%
Thus, only boosted radio signals could be detected for thermal branon with mass $M>6$ TeV (Right panel Figure \ref{Synchro_candidates}, dashed black line). 

Finally, we set constraints on the $\left(f, M\right)$ parameter space of extra-dimensional brane-world theories. 
In fact, when cosmological and astrophysical limits are imposed on brane-world models, i.e., the thermally averaged cross section is fixed, the tension $f$ becomes a function of the $M$ \cite{Cembranos:2003fu}, so that the mass of the DM particle remains as the only free parameter of the model. In the right panel of Figure \ref{Synchro_candidates} we show the constraints on the $\left(f, M\right)$ parameter space that could be obtained with the SKA1 telescope.
 There, the dashed black line represents the curve $f(M)$ associated to the thermal relic with $\left\langle \sigma v \right\rangle_{th}= 3\cdot10^{-26} \text{ cm}^{3}/\text{s}$. The region in red would be detectable by SKA1 without any boost.  This means that SKA1 could detect thermal branons with masses below 10 TeV and tensions $f<10$ TeV.  
Blue region in Figure \ref{Synchro_candidates} right panel represents the detectable parameter space considering a BH mass of $10^3 M_{\odot}$. 
Concluding, we consider two branon candidates: 1)  thermal branon ($f=27.5$  TeV, $M=50.6$ TeV) that well fits the gamma-ray cut-off observed by HESS at the Galactic Centre with $\langle \sigma v\rangle = 1.14 \cdot 10^{-26}$ ${\text{cm}^{3}/\text{s}}$ (yellow star in Figure \ref{Synchro_candidates}, right panel) \cite{Cembranos:2013pwa}; 2) a non-thermal branon ($f=4.99$ TeV, $M=38.1$ TeV) with averaged cross section of $\langle \sigma v\rangle = 1.76 \cdot 10^{-21}$ ${\text{cm}^{3}/\text{s}}$ 
fitting AMS-$02$ positron fraction excess \cite{Cembranos:2019amc} (pink star in Figure \ref{Synchro_candidates}, right panel).
%
%
\,Although some works justified such high boost \cite{ Bovy:2009zs}, it is unlikely that the branon models with such parameters provide the correct relic density within the standard freeze-out mechanism.

%

\section{Multiwavelength TeV DM}
\label{sec:Discussion}

\begin{figure*}
\centering
    \includegraphics[width=0.7\textwidth]{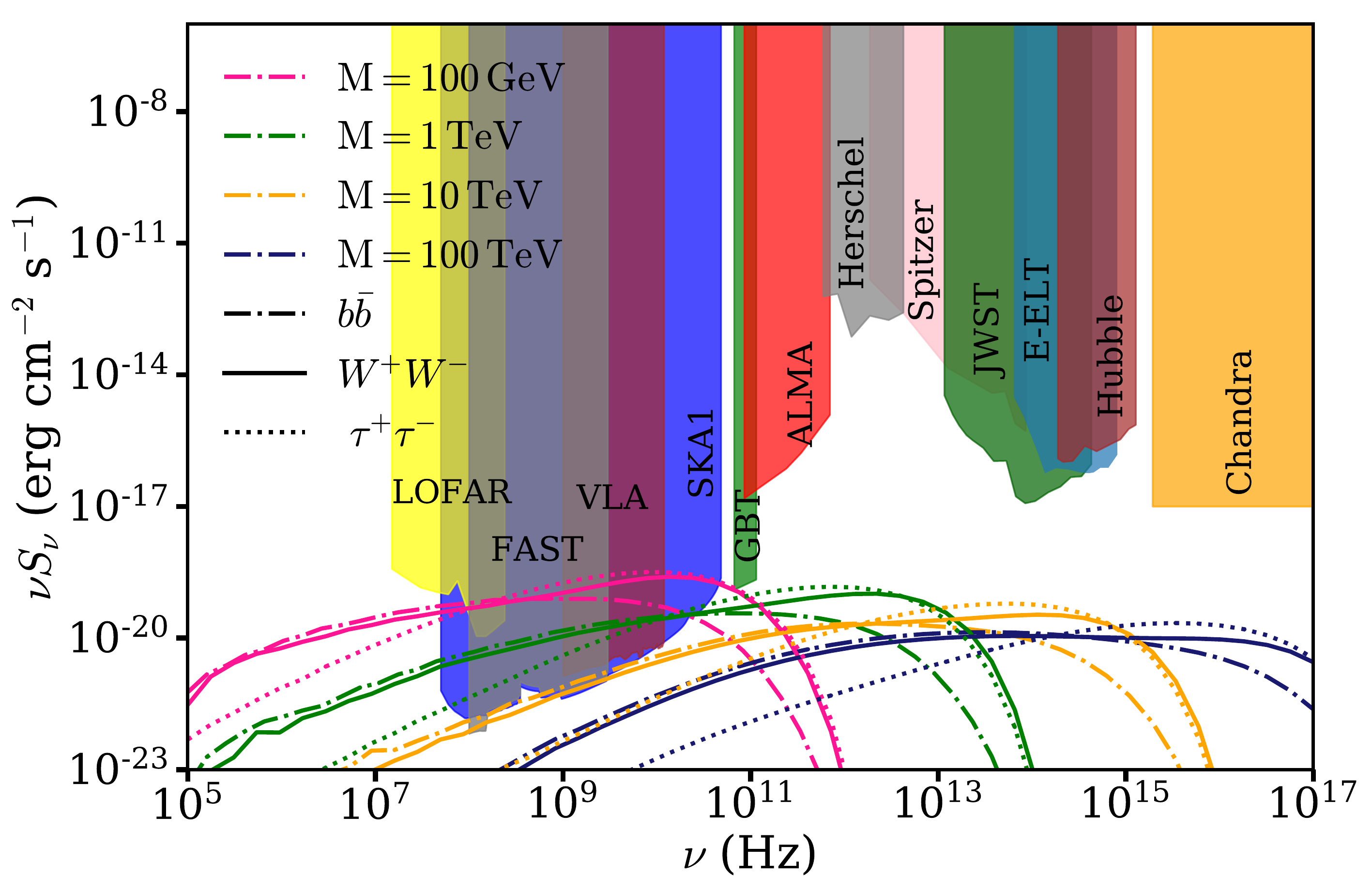}
\caption{
Flux density $S_{\nu}$  over $\Omega_{\text{DRACO}}$ as given in Eq. (\ref{S_syn}) times frequency $\nu$ for a Draco-like dSph in the range of frequencies $10^5-10^{17}$ Hz as produced by  $e^+e^-$ synchrotron emission for different DM masses and annihilation channels. 
Colour bands represent regions of detectability for different detectors: SKA1 \cite{Braun:2017hi}, FAST \cite{Dewdney:2013ard}, VLA \cite{Dewdney:2013ard}, LOFAR \cite{krankowski2014polfar}, GBT \cite{Dewdney:2013ard}, ALMA \cite{ALMAcalc}, Herschel \cite{ALMAcalc}, Spitzer \cite{ALMAcalc}, JWST \cite{JWSTsen}, E-ELT \cite{Dalcanton:2015jva}, Hubble \cite{ALMAcalc} and Chandra \cite{Chandrasens} on a large range of frequencies. 
Whereas GeV DM seems suitable to be detected in radio frequencies, TeV DM would be better detected at higher frequencies.  Even though SKA1 exhibits a competitive sensitivity to measure DM indirect signals up to 10 TeV (yellow lines), for heavier TeV DM, the maximum of emission shifts to frequencies higher than the SKA1 range. 
Targets with a radio signal boost mechanism could improve the competitiveness of these detectors.
 }
\label{Magnet1}
\end{figure*}

As thoroughly discussed in the previous sections, 
although SKA1 represents one of the most promising telescopes for radio astronomy and thus for multi-wavelength DM searches, the peak of synchrotron radio emission
from TeV DM  does not lie on the radio frequencies. In Figure \ref{Magnet1} we show how the higher the DM mass, the more the synchrotron emission peak is displaced to frequencies beyond radio, while the amplitude of the emission decreases with the mass. 
In such a figure, one can see how the peak of emission for DM masses of 1, 10, and 100 TeV lie on frequencies around $10^{12}$, $10^{14}$ and $10^{16}$ Hz respectively, quite away from radio frequencies and up to X-rays. 
However, in terms of sensitivity, SKA1 still remains as one of the most competitive
detectors according to Figure \ref{Magnet1}.
In this sense, prospective boosts due to a BH-induced DM-spike, could be important to detect synchrotron signatures beyond the radio frequencies with the current generation of experiments, such as  
%
ALMA, JWST and Chandra.


%
Concerning radio frequencies, 
we have mentioned in the introduction GBT, whose 
frequency band (67-115 GHz) 
is suitable to set constraints on TeV DM candidates with masses slightly heavier than what we can study with SKA, yet its sensitivity does not reach the detectability threshold for TeV DM candidates. Among the others, the  FAST and VLA telescopes should be considered: with their present designs, both of them are suitable for detection of DM masses lighter than $10$ TeV. However, the incorporation of new technology in both detectors is expected to improve their sensitivities and allow us to eventually detect heavier DM  \cite{2018SPIE10700E}. 
Finally, at low radio frequencies, LOFAR targets a frequency band ranging from 10 to 50 MHz (Figure \ref{Magnet1} in yellow) not achievable by other radio telescopes. Given the flux $\nu S_{\nu}$ dependence with the DM mass, we observe that LOFAR would be able to detect only tens GeV DM candidates. 
    \\


\section{Conclusions}
\label{sec:Conclusions}

In this work, we predict the prospective constraints we will be able to set on TeV DM candidates via the observation of the Draco dSph galaxy in radio frequencies with SKA. 

In the framework of radio telescopes, the specific intensity provides the angular distribution of the radio emission along the size of the source, while the flux density, refers to the integrated flux over either the angular size of the source or the beam selected for the analysis, as discussed in the text. In this sense, we can set constraints on a TeV DM candidate by studying the sensitivity of SKA with respect to the flux density expected by the annihilation of TeV DM particle in the selected source.
\\

\indent As shown in Figures \ref{Losses} and \ref{Magnet}, the crucial role that magnetic fields play in this kind of analyses should not be dismissed. For instance, the flux density times frequency for a 100 GeV DM candidate annihilating into $b\overline{b}$ channel provides a value of  $7.02\cdot10^{-20}\,\,\text{erg}\,\text{cm}^{-2}\,\text{s}^{-1}$ at $6\cdot10^{9}\,\,\text{Hz}$ when a magnetic field of $1\,\,\mu\text{G}$ is considered while, for the same frequency, the value becomes $2.42\cdot10^{-18}\,\,\text{erg}\,\text{cm}^{-2}\,\text{s}^{-1}$ if $B=5\,\,\mu\text{G}$. Thus an increment of just $4\,\,\mu\text{G}$ for the Draco magnetic field would enhance the radio signal around two orders of magnitude. In this regard, throughout our work, we used a conservative value of 1 $\mu$G for the magnetic field. We also show how the uncertainty related to the estimation of the diffusion coefficient affects both the intensity (Figure \ref{Magnet}) and the angular extension (Figure \ref{Losses}) of the radio emission. 
%

In Figure \ref{Synchro_SM}, we have compared the sensitivity of SKA with the synchrotron radio emission expected by the annihilation of model-independent thermal DM particles. 
We have concluded that SKA1-LOW optimizes the detection of sub-TeV DM, while the emission from heavier DM particles would be better detected by SKA1-MID.  
After $1000$ hours, SKA1 would make possible to detect DM candidates heavier than 10 TeV, provided DM annihilates mainly via either $t\overline{t}$ or $W^+W^-$ channels, while for DM annihilating via $\tau\overline{\tau}$ channel, only masses below 10 TeV would be detectable. 
In particular, in Figure \ref{Plot_sigmav} we set sensitivity constraints on the DM annihilation cross section and mass for model-independent DM candidates.  
%
%
We have also considered the case in which an intermediate-mass BH would be present in Draco, and we present the results of this study in Figure \ref{Synchro_BHplot}.
In this case, the flux density 
would be enhanced by more than three orders of magnitude, 
rendering very heavy DM candidates detectable by SKA1. 
%
In particular, $100$ TeV DM would be detectable within the 10-hours of data-taking.
\\

%
Finally, we consider extra-dimensional brane-world theories, i.e. branons as potential TeV WIMPs candidates. 
 %
%
%
%
In the left panel of Figure \ref{Synchro_candidates} we show the flux density expected from branon DM candidate for three  different masses we compare it with the sensitivity of the SKA telescope with three different observation time. We found that similar to the model-independent case, the SKA could constrain branon DM particles with masses less than 10 TeV and without any enhancement factor. On the other hand, in the right panel of Figure \ref{Synchro_candidates} we set constraints on the parameter space of branon candidates, i.e. the tension of the brane $f$ and the mass $M$ of the branon. In this plot, we show two branon DM candidates able to explain either the gamma rays cut-off detected by the HESS telescope (yellow star) or the AMS-02 excess in positron data (magenta star). This plot also shows the effect of a DM spike on constraining the mentioned parameter space.  

Concluding, we have discussed 
the multi-wavelength approach as a complementary strategy to detect TeV DM candidates. Although the electromagnetic radiation -  emitted by secondary products of TeV DM annihilation events via synchrotron emission  - cover frequencies up to the X-ray band, the current generation of detectors have no competitive sensitivity in order to set constraints on this TeV DM candidate, being SKA the more promising detector in this sense so far. The first SKA1 operations are expected to start in 2020 
and subsequently, with the addition of new technology, the progression to SKA2 will follow gradually  \cite{Dewdney:2013ard}, we expect SKA detectability regions to become even more competitive in the pertinent range of frequencies.
In this regard, we shall expect an improvement of a factor 10 improvement in the detection of radio signal (with respect to the one computed in this work) is expected whenever the SKA2 project will be fully operational.
\\
\\
\\
\\
{\it Authors’ note}: A preprint \cite{kar2019heavy} including a complementary model-
independent study was submitted to arXiv simultaneously to our communication by an independent group of authors. Unalike specific multi-TeV DM candidates are considered in each preprint.


\acknowledgments

This work was partly supported by the projects FIS2014-52837-P (Spanish MINECO) and FIS2016-78859-P (AEI / FEDER, UE).
AdlCD acknowledges financial support from projects FPA2014-53375-C2-1-P Spanish Ministry of Economy and Science, 
CA15117 CANTATA and CA16104 COST Actions EU Framework Programme Horizon 2020,  
CSIC I-LINK1019 Project, Spanish Ministry of Economy and Science,
University of Cape Town Launching Grants Programme and National Research Foundation grants 99077 2016-2018 (Ref. No. CSUR150628121624), 110966 Ref. No. BS1705-09230233 and the NRF Incentive Funding for Rated Researchers, Ref. No. IFR170131220846. AdlCD would like to thank the 
Abdus  Salam  International  Centre  for  Theoretical  Physics  (ICTP, Trieste) for its hospitality in the latest stages of the manuscript. 
VG's  contribution to this work has been supported by Juan de la Cierva-Formaci\'on FJCI-2016-29213 grant, the Spanish Agencia Estatal de Investigaci\'on through the grants FPA2015-65929-P (MINECO/FEDER, UE) and IFT Centro de Excelencia Severo Ochoa SEV-2016-0597, INFN project QGSKY, the Agencia Estatal de Investigaci\'on (AEI) and partially by the H2020 CSA Twinning project No.692194 ORBI-T-WINNINGO. VG thanks S. Camera, D. Gaggero, P. Ullio and  and the DAMASCO group for useful discussions and also acknowledges the support of the Spanish Red Consolider MultiDark FPA2017-90566-REDC. MMI would like to thank the SKA Pathfinder HI Survey Coordination Committee and the FAST-SKA Pathfinders Synergies meeting organising committee, specially to C. Carignan for its hospitality in first stages of the manuscript.
MMI thanks G. Beck, S. Blyth, J. Delhaize, M. Regis, H. Villarubia Rojo and S. White for useful discussions and acknowledges financial support from the University of Cape Town Doctoral Fellowships and the National Astrophysics and Space Science Programme (NASSP).

\bibliographystyle{unsrtnat}
\bibliography{skabibliography}
\end{document}